\documentclass[manuscript]{emulateapj}
\usepackage{graphicx}
\usepackage{amssymb}
\usepackage{epstopdf}
\usepackage{multirow}

\newcommand{\ha}{H$\alpha$}

\newcommand{\mic}{$\mu$m}
\newcommand{\bb}{$B_{\rm F475X}$}
\newcommand{\rr}{$R_{\rm F600LP}$}
\newcommand{\rrbis}{$R_{\rm F606W}$}
\newcommand{\ii}{$I_{\rm F814W}$}
\newcommand{\jj}{$J_{\rm F110W}$}
\newcommand{\hh}{$H_{\rm F160W}$}

\shorttitle{Stellar populations of quenched galaxies at $z \sim 1.5$}
\shortauthors{Bedregal et al.}

\begin{document} 
\title{HST/WFC3 Near-Infrared spectroscopy of quenched galaxies at {\it z} $\sim 1.5$ from the WISP Survey: Stellar population properties \altaffilmark{*}}

\author{A.G. Bedregal\altaffilmark{1,2,**}, C. Scarlata\altaffilmark{1}, A.\,L.~Henry\altaffilmark{3,4}, H. Atek\altaffilmark{5}, M. Rafelski\altaffilmark{6}, H.I. Teplitz\altaffilmark{6}, A. Dominguez\altaffilmark{7}, B. Siana\altaffilmark{7}, J.W. Colbert\altaffilmark{5}, M. Malkan\altaffilmark{8}, N.R. Ross\altaffilmark{8}, C.L. Martin\altaffilmark{3}, A. Dressler\altaffilmark{9}, C. Bridge\altaffilmark{9}, N.P. Hathi\altaffilmark{10}, D. Masters\altaffilmark{7,10}, P.J. McCarthy\altaffilmark{10}, M.J. Rutkowski\altaffilmark{1}}
\altaffiltext{*}{Based on observations with the NASA/ESA {\it Hubble Space Telescope}, obtained at the Space Telescope Science Institute, which is operated by AURA, Inc., under NASA contract NAS 5-26555.}
\altaffiltext{1}{Minnesota Institute for Astrophysics, University of Minnesota, Minneapolis MN 55455, USA}
\altaffiltext{2}{Department of Physics and Astronomy, Tufts University, Medford, MA 02155, USA}
\altaffiltext{3}{Department of Physics, University of California, Santa Barbara, CA 93106, USA}
\altaffiltext{4}{Astrophysics Science Division, Goddard Space Flight Center, Code 665, Greenbelt, MD 20771}
\altaffiltext{5}{Spitzer Science Center, Caltech, Pasadena, CA 91125, USA}
\altaffiltext{6}{Infrared Processing and Analysis Center, Caltech, Pasadena, CA 91125, USA}
\altaffiltext{7}{Department of Physics \& Astronomy, University of California Riverside, Riverside, CA 92521, USA}
\altaffiltext{8}{Department of Physics \& Astronomy, University of California Los Angeles, Los Angeles, CA 90095, USA}
\altaffiltext{9}{Department of Astronomy, Caltech, Pasadena, CA 91125, USA}
\altaffiltext{10}{Observatories of the Carnegie Institution for Science, Pasadena, CA 91101, USA}
\altaffiltext{**}{Contact email: {\it alejandro.bedregal@tufts.edu}}

\begin{abstract}
We combine HST G102 \& G141 NIR grism spectroscopy with HST/WFC3-UVIS, HST/WFC3-IR and {\it Spitzer}/IRAC[$\rm 3.6\mu m$]  photometry to assemble a sample of massive ($\rm log( M_{star}/M_{\odot}) \sim 11.0$) and quenched galaxies at $z \sim 1.5$. Our sample of 41 galaxies is the largest with G102+G141 NIR spectroscopy for quenched sources at these redshifts. In contrast to the local Universe, $z \sim 1.5$ quenched galaxies in the high-mass range have a wide range of stellar population properties. We find their SEDs are well fitted with exponentially decreasing SFHs, and short star-formation time-scales ($\rm \tau \le 100\,Myr$). Quenched galaxies also show a wide distribution in ages, between 1-4\,Gyr. In the $(u-r)_0$-versus-mass space quenched galaxies have a large spread in rest-frame color at a given mass. Most quenched galaxies populate the $z \sim 1.5$ red-sequence (RS), but an important fraction of them (32\%) have substantially bluer colors. Although with a large spread, we find that the quenched galaxies {\it on} the RS have older median ages (3.1\,Gyr) than the quenched galaxies {\it off} the RS (1.5Gyr). We also show that a rejuvenated SED cannot reproduce the observed stacked spectra of (the bluer) quenched galaxies {\it off} the RS. We derive the upper limit on the fraction of massive galaxies {\it on} the RS at $z \sim 1.5$ to be $<43\%$. We speculate that the young quenched galaxies {\it off} the RS are in a transition phase between vigorous star formation at $z>2$ and the $z \sim 1.5$ RS. According to their estimated ages, the time required for quenched galaxies {\it off} the RS to join their counterparts {\it on} the $z \sim 1.5$ RS is of the order of $\rm \sim 1\,Gyr$.

\end{abstract}
\maketitle

\section{Introduction}\label{sec:intro}

The formation and evolution of massive galaxies is one of the most studied  and debated topics in extragalactic astronomy today. In the local Universe the most massive galaxies primarily populate a well defined relation in the color-magnitude space known as the ``red-sequence'' \citep[e.g.,][]{kauffmann2003,baldry2004}. These galaxies host mostly old, passively evolving stellar populations \citep[e.g.,][and references therein]{trager1998, kuntschner2001, mehlert2003, sanchez-blazquez2006,renzini2006}.  The major processes behind mass assembly and structure formation of massive galaxies, however, are controversial.

 In particular, the role of major mergers \citep[as advocated in e.g.,][]{m-h1994,bekki1998,hopkins2006} in the mass build up and in the quenching of the star formation has recently been challenged by new evidence pointing towards a mass-induced truncation of the star-formation \citep[e.g.,][]{peng2010,peng2012}.
Evidence supporting the merger picture is mostly restricted to $z \lesssim 1$ from the tightness of the galaxy color-magnitude relation \citep[e.g.,][]{stanford1998, faber2007}, the evolution of the mass function of red-sequence galaxies \citep[e.g.,][; but see Cimatti et al.\,2007, Scarlata et al.\,2007]{bell2004, faber2007} and the small scatter in the mass-to-light ratios of these systems \citep[e.g.,][]{kelson2000}. The super-solar $\alpha$-element abundances found in local massive galaxies \citep[e.g.,][]{nelan2005, thomas2005} can be reproduced by both merger- and mass-induced quenching as long as the process took place several Gyr ago and on short time-scales \citep[$\rm \lesssim 1\,Gyr$,][]{thomas1999,renzini2009}.

Recently {\it minor} mergers have also been considered as a viable mechanism for the size growth of massive galaxies, gaining more popularity after the discovery of the strong size evolution of early-type galaxies \citep[e.g.,][]{bundy2009, deravel2009, lopez2010, ryan2012}. The discovery of the compact nature of massive quenched galaxies at $z \sim 2$ implies they must become $\sim 2$-$4 \times$ larger with time to match the sizes of quenched galaxies in the Local Universe \citep[e.g.,][]{daddi2005,trujillo2006,vandokkum2008,mclure2013}. Other authors, however, have suggested that the evolution of the mass-size relation is primarily driven by the appearance of new large galaxies at later time rather than by the growth of individual objects \citep[e.g.,][]{valentinuzzi2010,poggianti2012,carollo2013,cassata2013}.

The epoch between $1 \lesssim z \lesssim 3$ is crucial for the assembly of massive galaxies, as most of their size and number density evolution seem to have taken place during this time \citep[e.g.,][]{fontana2006,kriek2008,marchesini2009,ilbert2010,cassata2011,conselice2011}. Critical to study this redshift range are data covering the rest-frame optical, and in fact, the past few years have seen the completion of deep ground- and space-based photometric IR surveys.  Although large samples of both star-forming (SF) and quenched massive galaxies at $1 \lesssim z \lesssim 3$ can be assembled using color-selection techniques \citep[e.g.,][and references therein]{daddi2004, lin2012, newman2012}, results based on them are still hampered by the lack of spectroscopic redshifts, which translates into strong systematic effects on stellar population parameters (including, but not limited to, stellar masses and luminosity-weighted ages). Moreover, the mutual contamination between color-selected SF and quenched samples can only be established statistically as, typically, no further information is available for individual sources (e.g., $\sim 30\%$ contamination for $BzK$ selection, Cameron et al.\,2011). Detailed spectroscopic studies on individual sources are still extremely rare, and limited to the brightest, non representative sources \citep{kriek2009,vandokkum2010b,ferreras2012}. 

Additional complications appear when quiescent galaxies at $z > 1$ are considered. Without emission lines, the redshift determination must be constrained with continuum-emission features, like the 4000\AA\ and Balmer breaks, redshifted to the near-IR regime. This requires deep IR spectroscopic observations, which (until recently) were limited to single object spectroscopy.  These studies have shown that the bulk of the star formation in massive quenched galaxies took place between $2 < {\it z} < 4$ with formation time-scales below 1\,Gyr \citep{longhetti2005,kriek2006b,newman2010,onodera2012} . Ground-based near-IR spectroscopy, however, is limited to the $J$, $H$, and $K$ atmospheric windows, and suffers from high background emission and time-variable absorption features. Spectroscopic observations from space, then, represent the best way to acquire large samples of quiescent galaxies at $z \gtrsim 1$ with homogeneous spectroscopic data sets.

Here we present a stellar population study of a sample of quenched galaxies at $1.0 \leq  z \lesssim 2.0$ discovered in the  Wide Field Camera 3 (WFC3) Infrared Spectroscopic Parallel survey \citep[WISP,][]{atek2010}. The HST near-IR grism spectroscopy covers the wavelength range $0.9\mu$m $\leq \lambda \leq 1.6\mu$m, and allows us to study quenched high-$z$ galaxies on an individual basis.

The paper is structured as follows: in \S2 we present the observations and analysis of photometry and spectra; in \S3, we describe the color and magnitude selection of our sample. In \S4 we characterize our galaxy sample, including the calculation of spectro-photometric redshifts, stellar population parameters and estimates of their systematic and random uncertainties. In \S5 we discuss our main results, and summarize them in \S6. Five Appendices describing details in the data analysis are included at the end of the paper.

In this paper, we assume a flat cosmology with $H_0=\rm 70\,km\,s^{-1}\,Mpc^{-1}$, $\Omega_{\rm M}=0.3$, and $\Omega_{\Lambda}=0.7$. Photometric magnitudes are always expressed in the AB system \citep{okegunn1983}.

\section{Observations and Data}\label{sec:obs}

The sample presented in this work was extracted from the first 27 WISP fields
observed with the WFC3 on board of the
HST. Specifically, these fields were observed with both WFC3
grisms, and with both WFC3/IR and WFC3/UVIS cameras.  The WFC3/IR
provides a plate scale of 0.13$\rm ''/px$, over a total field of
view (FoV) of 123$''\times$136$''$.  The UVIS channel of the WFC3 has a
plate scale of 0.04$\rm ''/px$, over a total FoV of 162$"\times$162$"$.
For 7 of our fields the UVIS photometry was obtained from $2 \times 2$ binned data
(see Table\,\ref{tab:schedule}). All the observations were carried out in pure parallel
mode.\footnote{In the pure-parallel mode one or more instruments sample
  the HST focal plane while the prime program observes its planned
  target according to its desired visit schedule.}

In each field, imaging was obtained using the F475X, F600LP, F110W and
F160W filters, with typical exposure times of 400, 400, 1000, and 500
seconds, respectively (see Table\,\ref{tab:schedule} for
details). The two deeper fields Par96 and Par136 were observed in the
UVIS bands F606W and F814W instead. The WFC3/UVIS CCDs suffer a degradation of their charge
transfer efficiency with time, affecting F475X (F606W) and F600LP
(F814W) fluxes. A flux correction was implemented and it is fully
described in Appendix~\ref{app:cte}.

Dispersed images were obtained using the WFC3/IR camera and
the G102 and G141 near-IR grisms. The blue (G102) grism provides a
resolving power $R=210$ for a point source, and covers the
0.8-1.15\mic\ wavelength range.  The red (G141) grism provides a
resolving power $R=130$ for a point source, and covers the
1.07-1.7\mic\ range.  The wavelength overlap between the
two grisms ensures an accurate flux calibration of the spectra.  Details of reduction and calibration of the WFC3 data are presented in
\citet{atek2010}.

\begin{deluxetable*}{l c c c c c c c c c  }
\tabletypesize{\small}
\tablecolumns{9}
\tablewidth{0pt}
\tablecaption{Summary of Observations for the WISP Survey fields \label{tab:schedule}}
\tablehead{
\colhead{Field} & \colhead{   RA} & \colhead{DEC} & \colhead{F110W} & \colhead{G102} & \colhead{F160W} & \colhead{G141} &  \colhead{F475X} & \colhead{F600LP} & \colhead{IRAC\,3.6$\,\mu$m} \\
\colhead{ } & \colhead{[HMS]} & \colhead{[DMS]} & \colhead{[Sec]} & \colhead{[Sec]} & \colhead{[Sec]} & \colhead{[Sec]} & \colhead{[Sec]} & \colhead{[Sec]} & \colhead{[Sec]} \\
}
\startdata
Par95&       01 10 04.38  & $-$02 24 54.9    &   909   &5715   &384   &2209   &400   &400    &2100\\
Par91&       01 10 05.79  & $-$02 25 03.3   &    909   &5715   &384   &2209   &400   &400    &2100\\
Par97&       01 10 06.30  & $-$02 23 44.7    &   859   &5515   &406   &2109   &400   &400    &2100\\
Par84&       01 10 07.60  & $-$02 25 11.6    & 1162   &7521  & 559   &2809   &400   &400    &2100\\
Par79&       01 10 08.96  & $-$02 25 16.2   &  1187   &7521   &534   &2809   &400   &400    &2100\\
Par81&       01 10 09.12  & $-$02 22 17.1    & 1187  & 7521   &534   &2809   &400   &400    &2100\\
Par80&       01 10 09.27  & $-$02 22 17.7   &  1187   &7521   &534   &2809   &400   &400    &2100\\
Par89&       01 10 09.96  & $-$02 22 20.0   &  1187   &7512   &534   &2809   &400   &400    &2100\\
Par83&       01 10 10.71  & $-$02 24 09.7    &   859   &5515  & 406   &2109   &400   &400    &2100\\
Par96&       02 09 24.40  & $-$04 43 41.6    &   4295   &28081   &1765   &11430   &3000$^a$   &3000$^a$    &2100\\
Par74&       09 10 48.14  & +10 17 20.3    & 1065   &5918  & 431   &2306   &400   &400    &2100\\
Par87&       09 46 46.39  & +47 14 58.2    &   912   &4915  & 406   &1906   &400   &400    &1500\\
Par114 $^b$&     10 40 58.09  & +06 07 31.0    & 1137   &7221   &456   &2909   &600   &600    &---\\
Par131 $^b$&     10 48 22.94  & +13 03 50.5   &2171   &13039   &884   &5215   &600   &600    &---\\
Par115 $^b$&     11 18 55.08  & +02 17 09.6    &   912   &5215   &381   &2106   &600   &600    &---\\
Par135 $^b$&     11 22 24.01  & +57 50 58.9    &   862   &4712   & 406   &1906   &600   &600    &1500\\
Par136 $^b$&     12 26 28.84  & +05 23 02.9    &   3036   &18857   &1137   &7318   &2000$^a$   &2000$^a$    &---\\
Par76&       13 27 22.17  & +44 30 39.3    &   887   &5515   &406   &2006   &400   &400    &1500\\
Par120 $^b$&     13 56 51.50  & +17 02 33.9    &   837   &4512   &381   &1806   &600   &600    &1500\\
Par73&       14 05 12.86  & +46 59 19.9    & 1034   &6118  & 456   &2509   &400   &400    &1500\\
Par64&       14 37 29.04  & $-$01 49 49.5   &  1112   &5918   &456   &2306   &400   &400    &2100\\
Par66&       14 37 29.22  & $-$01 49 54.5   &  1237   &7421   &559   &2909   &400   &400    &2100\\
Par67&       15 24 07.75  & +09 54 53.9   &    959   &5715   &406   &2209   &400   &400    &1500\\
Par69&       15 24 09.75  & +09 54 50.0    & 1087   &5721   &431  & 2309   &400   &400    &1500\\
Par94&       22 05 26.66  & $-$00 17 48.5    & 1624  & 9024   &534   &3309   &400   &400    &2100\\
Par68&       23 33 33.04  & +39 21 20.5   &  1215   &7721   &534   &3009   &400   &400    &1500\\
Par147 $^b$&     23 58 19.72 &  $-$10 14 56.36 &   962&   5418&  406&   2106&  600&   600&  ---\\    
\enddata
\tablecomments{($^a$) UVIS observations of this field were acquired using the F606W and F814W filters. ($^b$) UVIS observations binned $2 \times 2$.}\\

\label{tab:obs_table}
\end{deluxetable*}

For 22 of the 27 fields we also obtained {\it Spitzer} IRAC
observations at 3.6\mic\ (see Table~\ref{tab:schedule} for
details). The same pipeline used to produce the  ``Spitzer Enhanced Imaging Products'' was also run 
on our data to measure the 3.6\mic\ fluxes  (Capak et al., in preparation).\footnote{For a full explanation
  of the pipeline see also http://irsa.ipac.caltech.edu/data/SPITZER/Enhanced/Imaging/
under ``Explanatory Supplement''.}

\subsection{Photometric Catalog}\label{sec:photcat}
In the analysis of the galaxy sample we simultaneously make use of the
broad band fluxes and near-IR spectra. Thus, special attention should be placed 
to make sure that the aperture used for the extraction
of the spectra and the aperture used to compute the broad band fluxes
are the same. In order not to introduce artificial correlations among
adjacent pixels we worked on images with the original pixel scales. As
a consequence, typical softwares used to compute object fluxes in
matched apertures in different images could not be used.

In order to create the multi-band catalog of aperture-matched fluxes we proceeded as follow. We first used the SExtractor software \citep{bertin1996} to detect objects in the deepest \jj-band images and to compute the parameters of the Kron elliptical apertures (semi-axes and position angle). The Kron apertures, as computed by SExtractor, are intended to give the most precise estimate of total magnitudes for galaxies. Details about the routine can be found in the SExtractor manual \footnote{http://www.astromatic.net/software/sextractor/ trunk/doc/sextractor.pdf}. For each galaxy, we then computed the total flux in all bands, using the \jj-band-defined aperture and a custom {\tt IDL} code.

The elliptical apertures were then appropriately scaled to account for different pixel scales between the UVIS and IR detectors. The local background for each elliptical aperture was calculated within a square annulus, using the same prescription as in SExtractor.  Within each annulus, the background per pixel was computed as a 3-$\sigma$ clipped mean.  Then, the total background-subtracted flux was computed within the aperture by adding all the aperture pixels.

Flux uncertainties were computed using Monte Carlo simulations. We created 1,000 images for each galaxy, by randomizing the flux in each pixel of the elliptical aperture according to a normal distribution with a width provided by the $1\,\sigma$ error map. We measure the total elliptical flux as explained above on each of the simulated images. We then computed the $1\,\sigma$ errors by fitting a Gaussian to the distribution of the 1,000 measured fluxes.  We only retain in the catalog sources with \jj-flux larger than $3\,\sigma$ (\jj\ is our deepest near-IR band, see Table~\ref{tab:schedule}). For undetected sources in the other bands, we show the $3\,\sigma$ flux limit in the figures.

Because of the significantly lower spatial resolution of the {\it Spitzer} images, IRAC 3.6\mic\ fluxes were computed independently from the HST fluxes, using the ``Spitzer Enhanced Imaging Products'' pipeline, with some modifications (details in Capak et al., in prep.). Briefly, the pipeline utilizes the Spitzer MOPEX software package to create a mosaic for each IRAC image, before identifying and extracting sources using SExtractor. Fixed aperture fluxes were computed for each object, using an aperture of 2\farcs8 diameter. This is a smaller aperture compered to the typical 3\farcs8 and 5\farcs8 diameter apertures used in the ``Spitzer Enhanced Imaging Products'' catalog.  However, our exposures are significantly deeper than the standard images analyzed with the pipeline, so a smaller aperture is more appropriate to reduce contamination from nearby neighbors. The aperture fluxes were corrected to total fluxes assuming the objects are point sources. This is a fair assumption as the FWHM of the {\it Spitzer} point-spread function (PSF from now on) of 1\farcs5 is larger than the FWHM in near-IR of the largest resolved sources used in this paper (1\farcs43).

The {\it Spitzer} and WFC3 catalogs were cross-matched using target coordinates, by searching within a radius of 1\farcs5. We flagged all sources for which multiple objects fell within the PSF of the IRAC data. This procedure results in a 5-broad-band photometric catalog for our 27 fields of 15,302 sources brighter than 27.5 mag in \jj-band.  Of those, 6,991 have IRAC 3.6\mic\ detections, 26\% of which are blended in this band.

\subsection{Near-IR Spectroscopy: 2D cleaning and 1D extractions}
Due to the slitless nature of the spectroscopic observations, spectra from different sources may overlap. If not properly removed, the flux contribution from nearby sources may substantially change the total flux and shape of the extracted spectra, preventing an accurate fit of the spectral energy distribution \citep[SED; e.g.,][]{gobat2013}. In order to perform an optimal cleaning of the 2D data \citep[reduced and callibrated with aXe software,][see Atek et al.\ 2010 for details]{kummel2009} the 1D spectra were extracted using a custom written {\tt IDL} code.  The detail of our extractions and a comparison with the aXe 1D extractions are presented in Appendix\,\ref{app:extraction}. Shortly, while aXe assumes a single Gaussian function to describe the spatial light profile of each source, we use a double Gaussian function. Our approach results in more reliable flux levels compared to extractions performed by the aXe code.

\begin{figure}[!t]
   \centering
   \includegraphics[width=8.5cm]{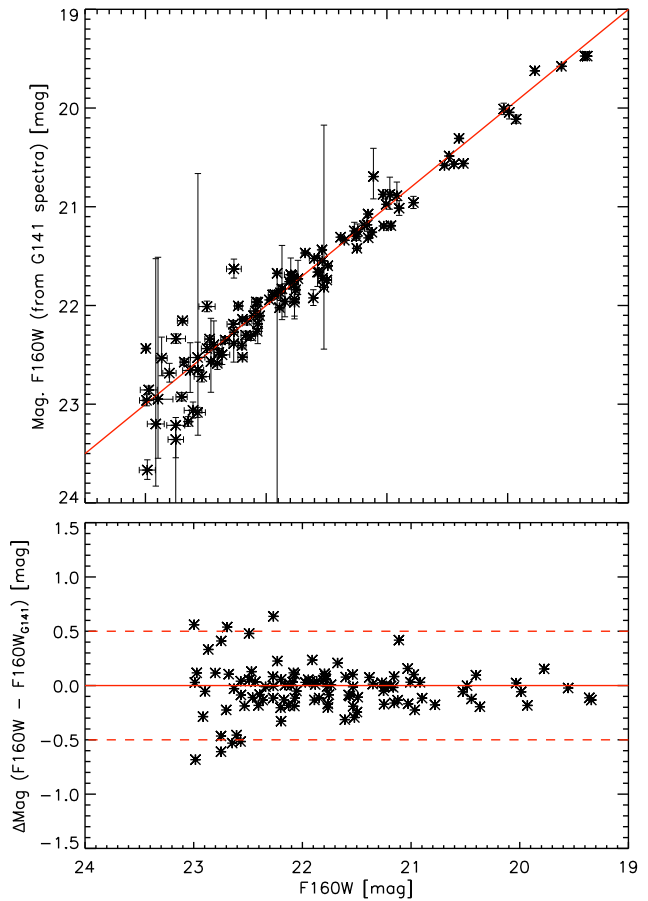}
   \caption{The comparison between the magnitudes computed from the extracted G141 spectra and those
     computed from the \hh\ images shows and overall good agreement between them. The comparison was based on 102 galaxies for which the spectral \hh\ magnitudes were calculated after contamination correction.}
   \label{fig:f160wg141comp}
\end{figure}

In order to test the quality of our spectral cleaning and extraction, we compare the fluxes from the spectra with our NIR photometric data.  In Figure~\ref{fig:f160wg141comp} we show a comparison between the magnitudes computed from the extracted G141 spectra and the magnitudes computed from the \hh\ images, for 102 galaxies. The magnitudes from the spectra were computed by convolving them with the throughput of the \hh\ filter, which is fully covered by the G141 spectral range. The median magnitude difference between photometry and grism data is $-0.01$\,mag with a $1\,\sigma$ scatter of $\pm 0.18$\,mag.

\section{Color-magnitude selection of the sample} \label{sec:selectETG} 
\begin{figure}[!h]
   \centering
   \includegraphics[width=9.0cm]{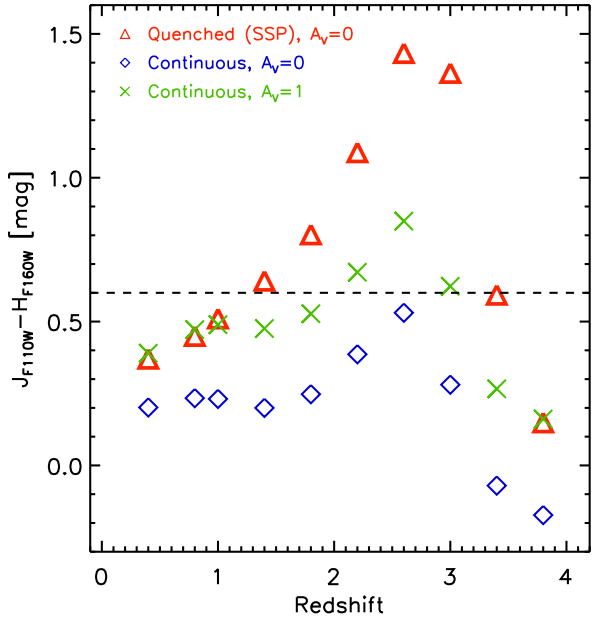}
\caption{The \jj$-$\hh\ color versus redshift diagram based on \citet{bc2003} models shows that with our color selection criterion (\jj$-$\hh $\geq$ 0.6, dashed black line) we mostly recover simple stellar population models (quenched, red triangles) between $1.2 \lesssim z \lesssim 3.0$. Dust-free continuous star formation models (blue diamonds) are rejected at all redshifts, while some dusty ($A_V=1$) continuous star formation models (green crosses) at $2.0 \lesssim z \lesssim 3.0$ pass our color cut. For all models, solar metallicity and a redshift of formation of 4 have been assumed.}

   \label{fig:color_z}
\end{figure}

The 4000\AA\ break fully falls within the wavelength range covered by the grism spectra for objects in the $1.2 \lesssim z \lesssim 2.7 $ redshift range. This break can be used not only as redshift indicator, but also to study stellar population properties in galaxies with sufficient S/N spectra \citep[e.g.,][]{ferreras2009, hathi2009,onodera2012}. We limit the analysis of the spectra only to objects with \jj$-$\hh\ $ \geq 0.6$, and magnitude brighter than \hh$=23$. As shown in Figure\,\ref{fig:color_z}, the color cut preselect sources with spectral breaks broadly covered by our spectra, for which the redshifts can be measured; while the magnitude cut ensures we have spectra with sufficient S/N.

Among all sources with \hh\ $\le 23$ we removed the stars using a size-magnitude diagram as explained in Appendix\,\ref{app:rm_stars}. After visual inspection of all the sources flagged as stars (315), we found that 10 were misclassified galaxies, and were re-included in the sample.

\begin{figure*}[!t]
   \centering
   \includegraphics[width=17.0cm]{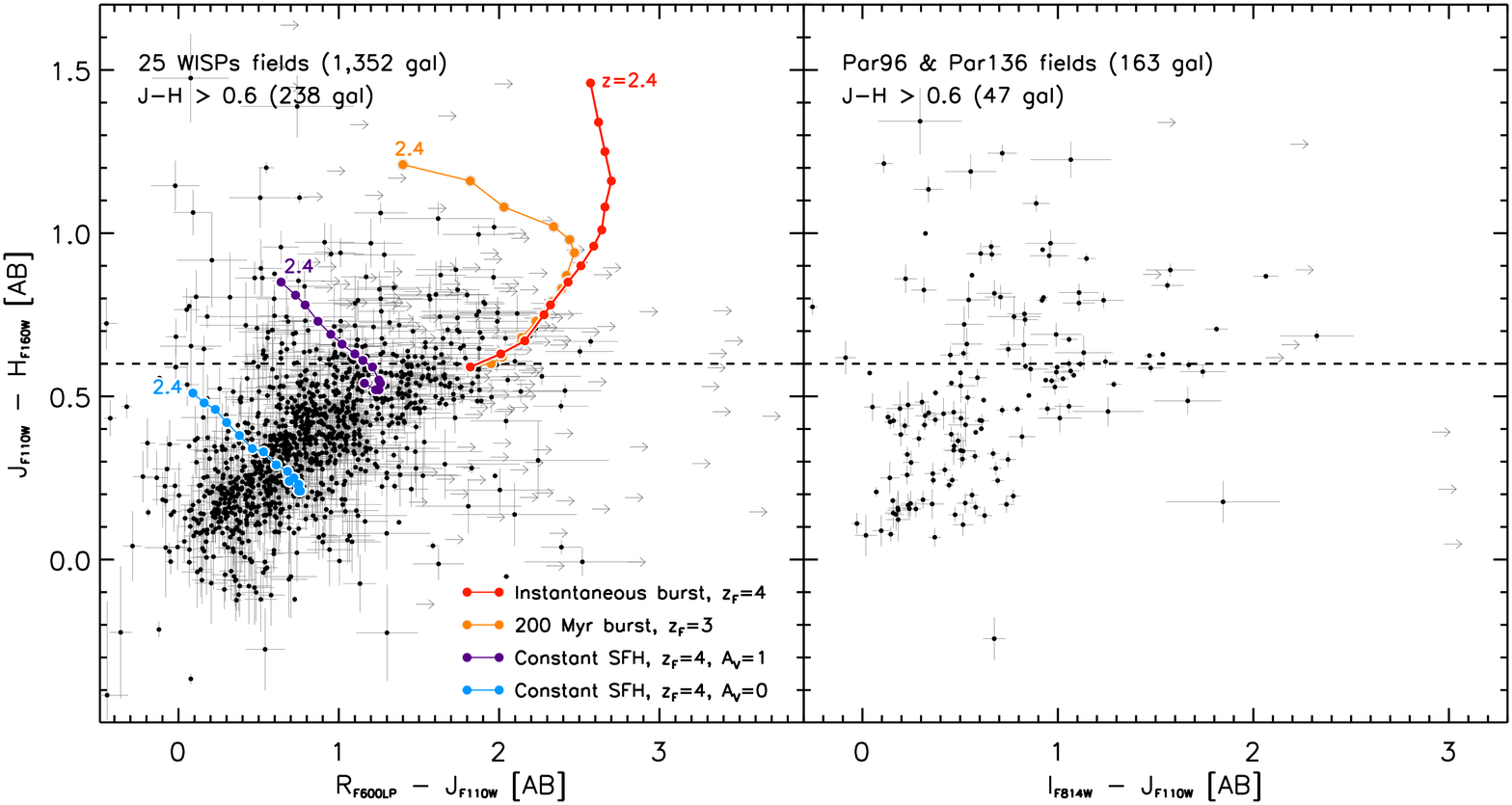}
\caption{The observed-frame color-color plots, showing galaxies from our photometric catalog with \jj\ $\leq 23$, illustrate how our color selection criterion (\jj$-$\hh $\geq$ 0.6, dashed black line) selects short-starburst galaxies (quenched, red and orange model tracks) and dusty continuous star-forming sources (purple model track). {\it Left}: We show data from 25 WISP fields using \rr$-$\jj. {\it Right}: We show data from the 2 deepest WISP fields using \ii$-$\jj\ color instead. Arrows indicate
  $3\,\sigma$ lower limits in \rr$-$\jj\ color, while dots indicate proper detections in all bands. The color tracks represent different star -formation histories from \citet{bc2003} models ranging from $1.0 \le {\it z} \le 2.4$ in steps of 0.1: light blue, continuous star formation; purple, continuous star formation assuming
  extinction $A_V=1$; red, single stellar population model (quenched); orange, starburst of 200\,Myr duration. For the starburst model a redshift of formation of 3 has been assumed. For all the other models a formation redshift of 4 was used. All models assume solar metallicity. The reddening vector changes as a function of redshift, and it is therefore not shown.}
  \label{fig:colselection}
\end{figure*}

In the left panel of Figure\,\ref{fig:colselection} we show the
\jj$-$\hh\ versus \rr$-$\jj\ color-color diagram for 1,352 galaxies
brighter than \hh$\,=23$. The right panel is the corresponding plot for
the two deep WISP fields, showing \jj$-$\hh\ versus \ii$-$\jj\ instead (163 galaxies).  
Sources brighter than \hh$\,=23$ constitute $\sim 11\%$ of our photometric catalogue. Out of 283
galaxies with \jj$-$\hh\ $\ge 0.6$, 84 (5) are upper limits in the \rr\ (\ii) band
and we show the $3\, \sigma$ limits in the color-color plots. 
Different model tracks showing the color evolution with redshift are overplotted (see figure caption and legend). Note that at the highest redshift considered ($z=2.4$), the truncated 200\,Myr burst model (orange line) is already passive, but significantly bluer in \rr$-$\jj\ than the instantaneous burst at the same redshift (red line). Figures\,\ref{fig:color_z} and \ref{fig:colselection} show that dusty SF galaxies
are also selected by our \jj$-$\hh\ color criterion. However, these
galaxies can be identified later through the SED fitting of their combined photometry$+$spectra
(see Section\,\ref{sec:results}).

We visually inspected both the spectra and the direct images of all the objects satisfying the \jj$-$\hh\ color cut to remove galaxies for which the spectrum showed instrumental issues (e.g., for sources at the edge of the chip for which the spectrum is truncated), for which the spectrum was contaminated at a level that even our procedure could not recover (e.g., close to a bright star), and for which the WFC3/{\it Spitzer} photometry show blended sources. We excluded 149 galaxies in this step. Note that about half of them were excluded because the full wavelength range of their spectra was not totally sample within the CCD detector (i.e., truncation). We also excluded 32 galaxies for which the 2D dispersed stamps did not show any signal. We compared these 32 galaxies with the parent sample, and found that they are at the faint end of the considered magnitude range (with average \hh-band of 22.7), and tend to have slightly larger radii than the remaining sources of the same magnitude. This may introduce a bias against less concentrated objects in our faintest magnitude bin.

Our {\it master} sample includes 102 sources with \jj$-$\hh $\geq 0.6$ and \hh $\le 23$. In 23 out of 102 galaxies we detected emission lines. It is important to note that, because in slitless mode the spectral resolution depends on the size of the object in the dispersion direction, our capability to detect emission line galaxies depends not only on the line flux and equivalent-width, but also on the size of the galaxy itself. A detailed line-completeness analysis for the full WISP survey is presented in Colbert et al.\,(2013, submitted). The emission line recovery rate drops below 40\% for objects larger than 0\farcs6, The median size of our sample galaxies is 0\farcs45, corresponding to a completeness of about 70\%. As discussed in Colbert et al., the completeness never reaches 100\% because of spectral contamination issues. Our sample of 102 galaxies does not suffer from this problem, so we expect the completeness rate in our case to be even larger than 70\%.

\section{Analysis}

Previous to derive spectro-photometric redshifts and stellar population properties (SPPs) from our analysis, we study the effects of emission lines in retrieving these continuum-based parameters. Our low spectral resolution might ``dilute'' emission lines in the continuum, affecting the retrieved redshifts and SPPs. The results of this study are presented in Appendix\,\ref{app:emilines}. Briefly, we find that in those cases were emission lines are diluted in the continuum, the flux contribution from the line is below the $\rm 1\,\sigma$ uncertainties in the grism data. Therefore, in addition of masking the detected emission lines in the SEDs, there is no need to make specific modeling of emission lines for our SED fitting process and simulations.

\subsection{Spectro-photometric redshift}\label{sec:redshift} 
We compute spectro-photometric redshift for the 102 galaxies in our sample by fitting stellar population models (see below) to the combination of the photometric and spectroscopic data. We developed our own {\tt IDL} code for this task because of the $1)$ different spectral resolution in the two grisms, $2)$ different spectral resolution in each individual source, and $3)$ simultaneous fit to photometry and spectra.

We computed spectro-photometric redshifts by fitting 51 empirical templates from \citet{cww1980} and \citet{kinney1996} to the \bb (\rrbis), \rr (\ii), G102 and G141 spectra, and IRAC 3.6\mic\ data, when available. Before the $\chi^2$ minimization, we matched the spectral resolution of the templates to that of the data (on an object-by-object basis). To do so, we re-binned both the data and the templates to the resolution element, computed from the FWHM of the object's spatial profile in the dispersion direction. Given the typical size of our sources, the spectral resolution elements are on average $\sim 85.8$\AA\ ($\rm 162.8\AA$) in G102 (G141).  We considered the $0.0\leq z \leq 4.0$ range with constant redshift steps of 0.05, and applied intergalactic medium absorption following \citet{madau1995}.   For sources with emission lines the purely spectroscopic redshift based on those features was used instead ($19\%$ of our master sample).  We determined redshifts for single-line galaxies assuming they are H$\alpha$ emission. See \citet{dominguez2012} and Colbert et al.\,(2013, submitted) for further details on emission line redshift estimations.

We estimated random errors on the redshifts by using Monte Carlo simulations. For each galaxy, we varied both the spectra and the photometric points within the $1\sigma$ uncertainties, assuming Gaussian distribution. We created 500 realizations for each galaxy, and derived the random error on the redshift as the 1\,$\sigma$ width of the resulting distribution.  

\begin{figure}[!h]
   \centering
  \includegraphics[width=9.0cm]{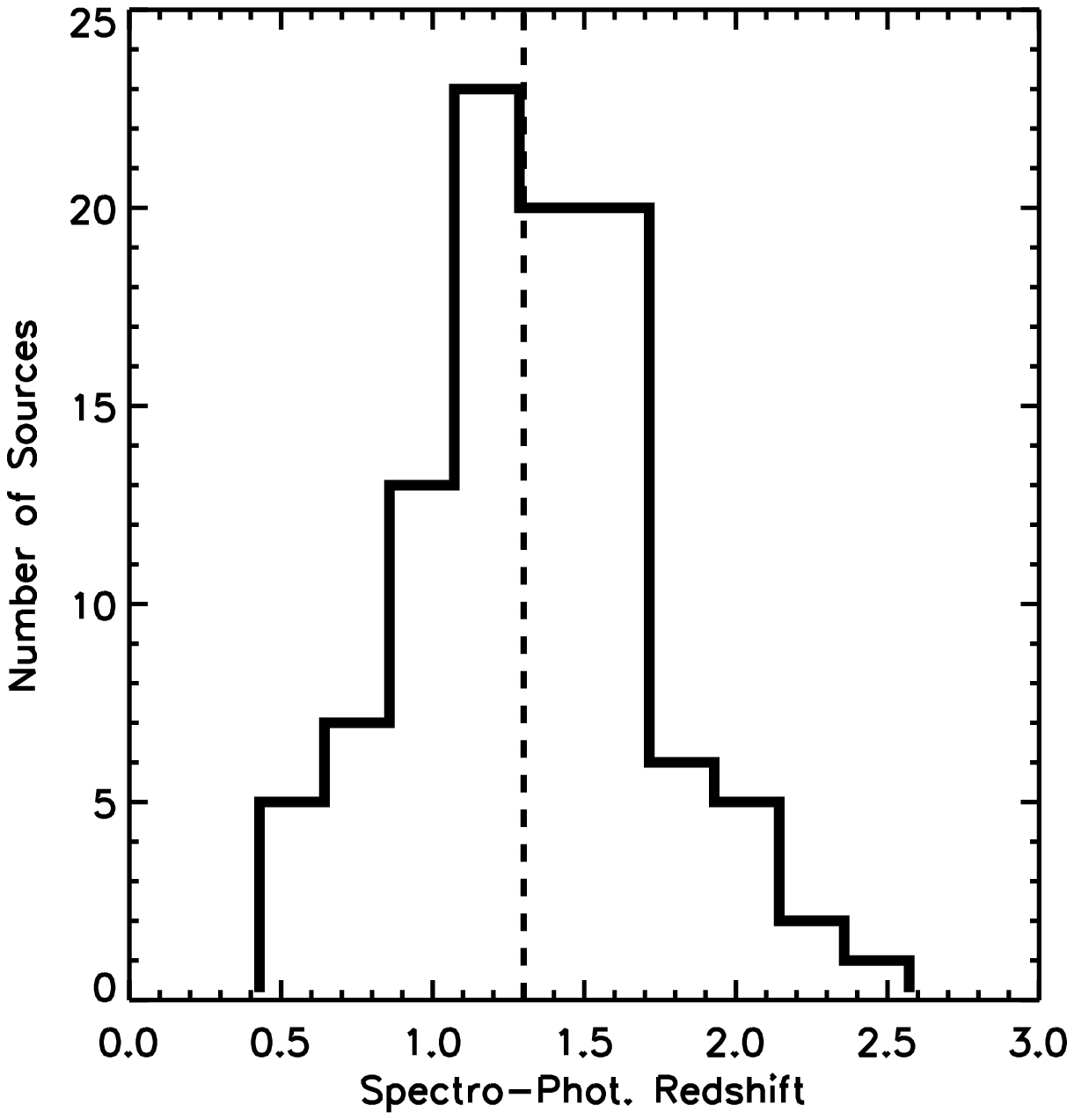}
 \caption{The spectro-photometric redshift distribution for the master sample of
   102 galaxies shows a median value of $z=1.3$ (dashed line).} 
  \label{fig:z}
\end{figure}

\begin{figure*}[!t]
   \centering
  \includegraphics[width=18.0cm]{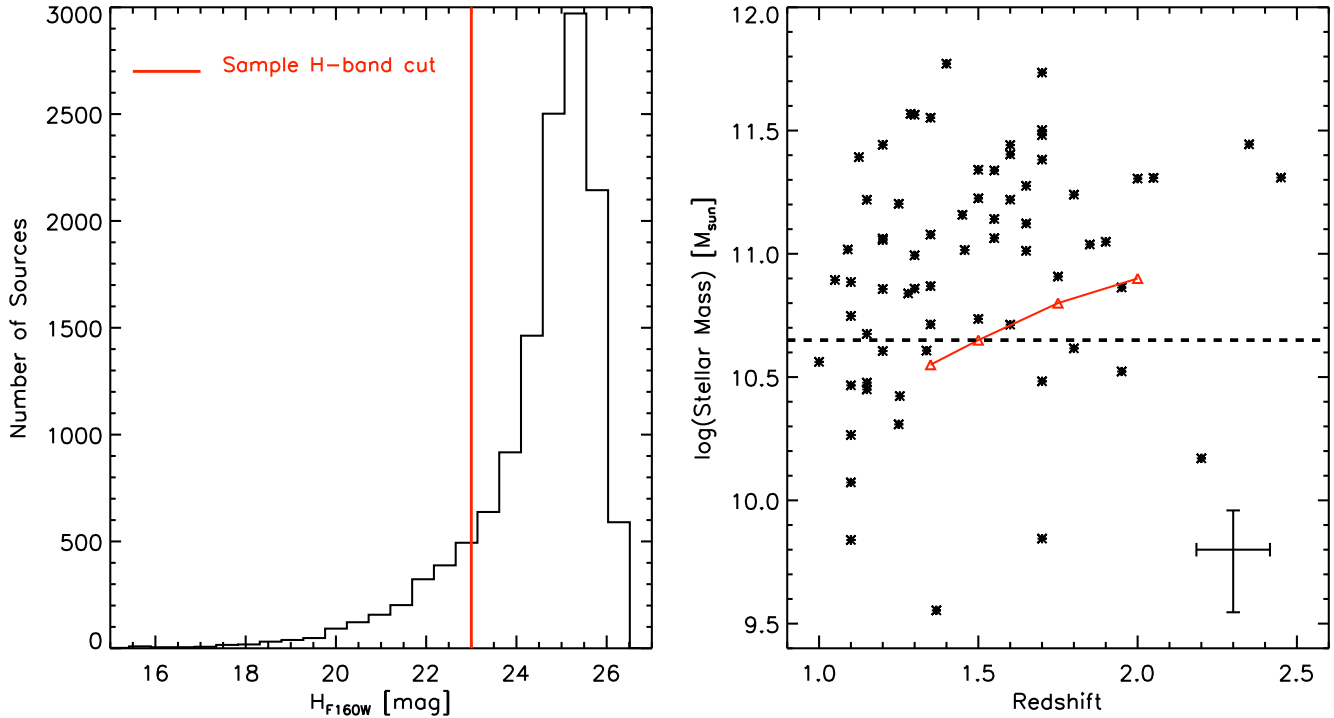}
 \caption{{\it Left}: The \hh-band magnitude histogram for our entire photometric catalog shows it is highly complete above our \hh$=23.0$ magnitude cut (red line). {\it Right}: In the stellar mass versus redshift diagram for our sample galaxies (after removing cases with potentially strong systematics, see Section\,\ref{sec:systerr}) we overplot \citet{bc2003} model predictions for $100\%$ mass completeness at different redshifts (red line) and our mass completeness cut at $\rm log(M_{star}/M_{\odot}) = 10.65$ (black dashed line). Error bars in the bottom-right corner represent median total uncertainties (random and systematic).}
  \label{fig:masscompl}
\end{figure*}

The spectro-photometric redshift distribution for the full master sample of
102 galaxies is shown in Figure~\ref{fig:z}. The distribution shows that 77\% of the
selected galaxies are at $z>1$.

\begin{table*}[!t]
\begin{center}
\caption{Systemmatic errors in SPPs $^*$\label{tab:systerr}}
\begin{tabular}{|l|cr|cr|}
\hline
Param. & Syst.Error (with IRAC) & & Syst.Error (no IRAC) & \\
\hline
\multicolumn{1}{|c|}{\multirow{3}{*}{Redsh}} & $1.0 \leq z \leq 1.2:$ &  $0.0 \pm 0.10$      &  $ $  &\\
                                                                      & $z >1.2$                    &  $0.0 \pm 0.05$       &  $ $  &\\
\hline

\multicolumn{1}{|c|}{\multirow{2}{*}{Age $^a$}} & $\rm 1.3 \leq {\it z} \leq 1.8$ \& $\rm Age \leq 1.5\,Gyr:$ &  $ 0.0 \pm 75\%$      & $\rm 1.3 \leq {\it z} \leq 1.8$ \& $\rm Age \leq 1.5\,Gyr:$ &  $ 0.0 \pm 65\%$ \\
                                                                           & $\rm {\it z} > 1.3$ \& $\rm Age > 1.5\,Gyr: $                     &  $0.0 \pm  35\%$      & $\rm {\it z} > 1.3$ \& $\rm Age > 1.5\,Gyr: $                      &  $ 0.0 \pm 45\%$ \\
\hline

\multicolumn{1}{|c|}{\multirow{2}{*}{M$_{\rm star}$$^b$}} & $z < 1.2: $ &  $ \approx 20\%$              &   &  \\
                                                                                         & $1.2 \leq z < 1.8: $ & $10\%$ &   &   \\
                                                                                         & $z \geq 1.8: $ &  $10\%$          &  $z \geq 1.8: $ &  $15\%$ \\

\hline
\multicolumn{1}{|c|}{\multirow{5}{*}{SFR $^c$}} & $\rm SFR < 2.5\,M_{\odot} yr^{-1}:$ &  $\rm 0$                                                 &  &   \\
                                                                          & $\rm  2.5 \leq SFR \leq 7.5\,M_{\odot} yr^{-1}: $ &  $\rm  0.0 \pm 1\,M_{\odot} yr^{-1}$ & $\rm  2.5 \leq SFR \leq 7.5\,M_{\odot} yr^{-1}: $ &  $\rm  0.0 \pm 2\,M_{\odot} yr^{-1}$ \\
                                                                          & $\rm  7.5 \leq SFR \leq 15\,M_{\odot} yr^{-1}: $ &  $\rm  0.0 \pm 3\,M_{\odot} yr^{-1}$     & $\rm  7.5 \leq SFR \leq 15\,M_{\odot} yr^{-1}: $ &  $\rm  0.0 \pm 5\,M_{\odot} yr^{-1}$ \\
                                                                          & $\rm  15 \leq SFR \leq 20\,M_{\odot} yr^{-1}: $ &  $\rm  0.0 \pm 8\,M_{\odot} yr^{-1}$      & $\rm  15 \leq SFR \leq 20\,M_{\odot} yr^{-1}: $ &  $\rm  0.0 \pm 15\,M_{\odot} yr^{-1}$ \\
                                                                          & $\rm  SFR > 20\,M_{\odot} yr^{-1}: $ &  $\rm  25 \pm 20\,M_{\odot} yr^{-1}$                    & $\rm  SFR > 20\,M_{\odot} yr^{-1}: $ &  $\rm  45 \pm 40\,M_{\odot} yr^{-1}$ \\

\hline
\end{tabular}
\end{center}
\footnotesize{($^*$) In the 'no IRAC' column, if an empty space is present, the results are the same as in the 'IRAC' case. ($^a$) Only ages for $z\geq 1.3$ galaxies are considered in this work. ($^b$) At $z < 1.2$ the systematic errors in stellar mass vary monotonically from 30, 20 and 10\% at $\rm log M_{star}/M_{\odot} = 11.8$, 11.0 and 10.6, respectively. ($^c$) At ${\it z} \geq 1.3$ the error is valid for $\rm SFR < 20\,M_{\odot} yr^{-1}$ only.}
\end{table*}

\subsection{Stellar population properties}\label{sec:sedfit}
We compute luminosity-weighted age, stellar mass, star
formation rate (SFR) and star-formation history (SFH) of the master sample galaxies,
using the same custom {\tt IDL} code applied in the previous section, keeping the redshift fixed and using the
\citet{bc2003} library of stellar population synthesis models. We
consider seven SFHs (continuous, exponentially declining with
{\it e}-folding times $\rm \tau=0.01, 0.1, 1, 5\,$Gyr, and exponentially
increasing with $\rm \tau=1, 5\,$Gyr); 70 log-binned ages between
10\,Myr and 12\,Gyr; Salpeter initial mass function \citep[IMF,][]{salpeter1955} and solar metallicity. 

We use a Salpeter IMF based on recent results on local massive galaxies. A variety of observations,  including stellar kinematics \citep[e.g.,][]{dutton2012,cappellari2012}, stellar populations \citep[e.g.,][]{conroy2012,spiniello2012,tortora2012} and gravitational lensing \citep[e.g.,][]{auger2010,treu2010,brewer2012}, suggest ellipticals and spiral bulges show ``heavier'' IMFs (Salpeter-like, with larger fractions of low-mass stars) than galaxy disks \citep[Chabrier-like IMF,][]{chabrier2003}. Thus, the assumption of a Salpeter IMF seems to be more representative of the massive and quenched galaxy population we address in the present study. We note that if a Chabrier-like IMF was used, our inferred masses and SFRs would be smaller by a factor $\approx 1.7$. Accordingly, our models and the mass-completeness limit we inferred from them (see Section~\ref{sec:masscompl}) would be smaller in a similar amount without compromising our main results and conclusions. We also note that our {\it specific} SFRs (SSFRs) are robust to IMF changes. This is of particular importance as we base our quenched/SF classification on this parameter. Systematic changes of the IMF with galaxy mass are unlikely in the mass range considered in this work, and are briefly discussed at the end of Appendix~\ref{app:systerr}.

Solar metallicity is in agreement with recent spectroscopic results by \cite{onodera2012} on quenched galaxies at $z \sim 1.4$. We considered a range of extinctions ($0\leq A_V \leq 1$), and we used the \citet{calzetti2000} extinction law. The age of the stellar population is constrained to be smaller than the age of the Universe in the adopted cosmology.  Analogously to the redshift random error determination, random uncertainties on the SPPs were derived using Monte Carlo simulations.

We used the results of the stellar population modeling to select quenched galaxies as commonly done in the literature, based on the SSFR: {\it quenched} galaxies have $\rm SSFR < 0.01\,Gyr^{-1}$, and constitute 71\% of our master sample.

\subsection{Systematic uncertainties in redshift and stellar population parameters}\label{sec:systerr}
The statistical uncertainties associated with the measurements of redshifts and stellar population parameters are assessed in Sections~\ref{sec:redshift} and \ref{sec:sedfit}. Here, we further quantify the {\it systematic} uncertainties introduced by the degeneracies in stellar population models.  It is well known that, if the photometric/spectroscopic data do not provide adequate constraints, studying SPPs of galaxies through SED fitting may produce strongly degenerate results by different combinations of age, SFR, SFH and extinction, \citep[e.g.,][and references therein]{muzzin2009}.  For this reason we performed a set of simulations to assess how well we can recover the SPPs with the available data. 

The simulations are described in detail in Appendix\,\ref{app:systerr}. Briefly, we used \citet{bc2003} models with known stellar population properties and redshifts to simulate our spectroscopic and photometric data. These ``model data'' were treated in the same way as the real data: we applied both the color and magnitude cuts as for our galaxies, and the selected model data were fitted with our customized {\tt IDL} code to recover their SPPs. In this section, we only summarize the main results of these tests and their implications for our galaxy sample.

Not surprisingly, the main conclusion of our study was that, in general,  the SPPs were recovered more accurately for galaxies at $z\gtrsim 1.3$, where the 4000$\, \rm \AA$ break was fully covered by the grism spectra. More degeneracies appeared when the break was only coarsely covered by the UVIS photometric points.

At redshift $z \geq 1.3$, our simulations show we can recover the age measurements to within $\sim 35\%$ of the age value for our complete age range. We can also distinguish between short ($\tau \leq 100$\,Myr) and long ($\tau \geq 1$\,Gyr) SFHs, although we can not separate SFHs with $\tau = 100$\,Myr and $\tau = 10$\,Myr. Most importantly, we find that our distinction between quenched and SF galaxies (defined by a rough limit at  $\rm SSFR= 10^{-2}\,Gyr^{-1}$) is very robust.

At lower redshifts, we found we cannot reliably recover spectrophotometric redshifts below 1. Therefore we decided to exclude from our galaxy sample all sources below this redshift. At $1<z<1.3$ we recovered most of the SPP values as at $z>1.3$, although with larger scatter. At these lower redshifts, however, we found that the age and extinction show strong systematic offsets with respect to the input parameter values. This was mainly produced by a strong anti-correlation between both galaxy properties. However, because the distinction between quenched and SF galaxies is defined as function of the SSFR only, even at $z<1.3 $ we can robustly distinguish between both galaxy types. The stellar mass is the most robust and best constrained parameter at all redshift. 

Overall, the addition of IRAC $\rm 3.6\,\mu m$ data to the SED fits slightly reduces the scatter on the ages and stellar-mass-systematic uncertainties. The use of IRAC data provides the strongest constraints in the determination of SFR, were the scatter of our models is clearly reduced. The recovered redshifts also suffer a very mild improvement in their systematic uncertainties when IRAC data is used, while the other SPPs are mostly unchanged.

From the discussion above, we decided to remove from the following analysis all galaxies with $z < 1$. Also, we will not to use in our analysis the age of galaxies at $z<1.3$ and the extinction of all galaxies independently of their redshift. 

In Table\,\ref{tab:systerr} we present a summary of the results from the systematic error analysis for redshift and SPPs, including results with and without IRAC data in the SED fits. Formally, many of the systematic offsets found are consistent with zero within the uncertainties. However, some cases have larger uncertainties than others. Therefore in those cases we present our results as zero-systematic-error $\pm$ their 68\%-percentile.

\subsection{Stellar Mass completeness}\label{sec:masscompl}
The estimate of the mass completeness is not straight forward in a magnitude {\it and} color selected sample like ours. Together, these criteria imply not only a segregation in mass but also in current star formation activity and SFH.  

First we explore the \hh-band completeness of our photometric catalog. In the left panel of Figure\,\ref{fig:masscompl} we show the \hh-band histogram for our entire photometry (Section\,\ref{sec:photcat}). We proceed by run 1,000 simulations of our \hh-band FoV and re-extract with SExtractor our simulated galaxies. Each of the 1,000 galaxies per simulation was modeled as a random combination of observational properties which ranges were set based on the data. We allowed a range of \hh-band magnitudes between 26 and 15. Then, light profiles were modeled with a 2D S$\rm \grave{e}$rsic function with $0 \leq n \leq 4$, centered at random positions, with ellipticities between 0 and 1, and random projected orientations. The effective radii were allowed to vary between 0.2 and 0\farcs6 (typical range for our galaxies). We add random noise to every single model run such as an average \hh=22 mag galaxy has an integrated S/N=200, like in the data. After extracting galaxies from each simulation we study the completeness as a function of \hh-band magnitude, in bins of 0.2\,mag. We found we are 100\% complete down to magnitudes of 24.2., reaching 50\% completeness at 25 mag. We also study completeness as function of other parameters. Based on the effective radius we are 100\% complete up to 0.35 arcseconds, reaching 70\% completeness at 0\farcs43. In terms of the S$\rm \grave{e}$rsic index we are 70\% complete down to $n=2$. Finally, we are 75\% complete for ellipticities up to 0.4. Based on these results, our magnitude cut at \hh$=23$ for our galaxy sample (Section~\ref{sec:selectETG}) is conservative enough to allowed us study a 100\% complete sample in \hh-band with also very high completeness in other observables.

In the right panel of Figure~\ref{fig:masscompl} we show the galaxy stellar mass as a function of redshift for our galaxies (after cleaning the sample of cases with possible strong systematics in their SPPs, see previous section). Due to our \hh$=23$ magnitude cut, the lower mass envelope changes with redshift, and depends on the galaxy SFH and age. In order not to introduce any bias in the analysis, we conservatively estimate the mass completeness limit to be the mass of the model with the highest M/L ratio for \hh$=23$ at each redshift, as indicated by the solid red curve in the figure\footnote{The red curve shows the
  stellar mass corresponding to \hh$=23$, for an exponentially
  declining SFH, with $\tau=10$ Myr, and formation redshift $z=5$}.  In what follows, we adopt a mass completeness limit of $4.5\times 10^{10}$~M$_{\odot}$ (dashed black line in right panel of Figure~\ref{fig:masscompl}), which is the minimum mass measurable at ${\it z} \sim 1.5$, with our conservative assumptions. We note that $\sim 70\%$ of our galaxies were above this mass cut. Of them,  40\%  are at or below redshift 1.5.

\subsection{The final sample of quenched galaxies}

\begin{figure*}[!p]
   \centering
  \includegraphics[width=17cm]{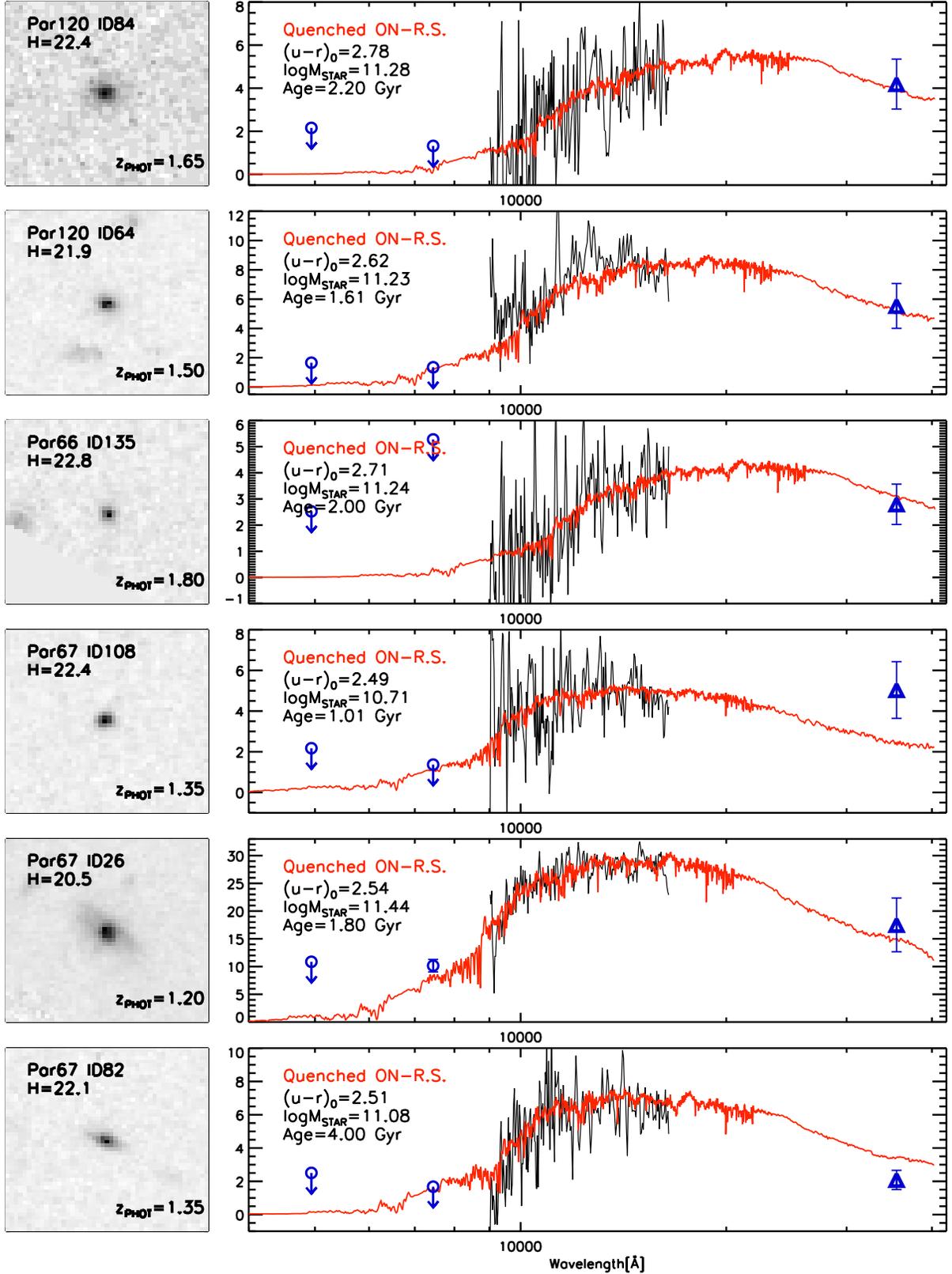}
  \caption{Examples of SED fits for our sample galaxies (fits for our 41 galaxy sample available in the electronic format of the paper). In the left column $6 \times 6$ arcsec$^2$ image stamps in \hh-band. In the right side,  the data and SED fits. In black, the G102 and G141 spectroscopic data. In blue, the photometric data. Circles represent HST/WFC3/UVIS \bb\ and \rr\ photometry (arrows correspond to $3\,\sigma$ upper limits). Triangles represent {\it Spitzer}/IRAC\,$3.6 \mu$m observations (when available). In red the best SED model fit from our library of \citet{bc2003} stellar population models. In the Y-axis, flux densities, $\rm F_{\lambda}$, are normalized to $\rm 10^{-19} erg s^{-1} cm^{-1} \AA^{-1}$.}
  \label{fig:spectra}
\end{figure*}
 
The final sample of quenched galaxies includes sources at $1.0 \leq z \lesssim 2.0$. At $ z >2.0$ our mass limit implies that we find only a few of the rarest most massive galaxies, while at $z<1.0$, our simulations show that we cannot reliably retrieve photometric redshifts with our data (see Section\,\ref{sec:systerr} and Appendix\,\ref{app:systerr}).

After the redshift and SSFR cuts, our sample includes 41 quenched sources with $H \le 23$, $J-H \ge 0.6$, $\rm log(M_{star}/M_{\odot}) \geq 10.65$, and redshifts in the range $1.0\leq z \lesssim 2.0$. Four of the quenched galaxies ($10\%$) have detected emission lines.

  We do not remove these galaxies from the sample of quenched objects for the following reasons: first, if star-formation is at the origin of the detected emission lines, the line fluxes implies SFRs of about  0.4\,$\rm M_{\odot} yr^{-1}$ (using a Kennicutt 1998 conversion), which, given the masses of these objects, would still make them quenched according to our SSFR definition. Second, the emission lines might be produced by ionization due to an active galactic nucleus (AGN), rather than star formation.  Various studies connect AGN activity with the quenching of star formation \citep[e.g.,][]{dimatteo2005, hopkins2006,oser2012}, so we do not want to bias our result against AGN activity. Because we lack information on the origin of the gas ionization, we leave the 4 emission line galaxies in the sample, and identify them in the analysis when needed.  As it will be  shown in the following sections, the presence of quiescent galaxies with emission lines does not affect our main results and conclusions.

In Figure\,\ref{fig:spectra} (available in the electronic format of the paper) we show the \jj-band postage stamp of each galaxy of our final sample. In the same figure we present the G102 and G141 spectra together with the WFC3/UVIS and IRAC 3.6\mic\ photometric points used in the SED fits.  The best SED model fit and some SPPs are also included. This 41-galaxy sample will be used in the rest of the paper.

\section{Results and discussion}\label{sec:results}

In the following section we discuss the properties of the
final sample of 41 quenched galaxies.

\subsection{Distributions of stellar population properties of the quenched galaxy sample}\label{results:sampleSPP}

\begin{figure}[]
   \centering
   \includegraphics[width=8.0cm]{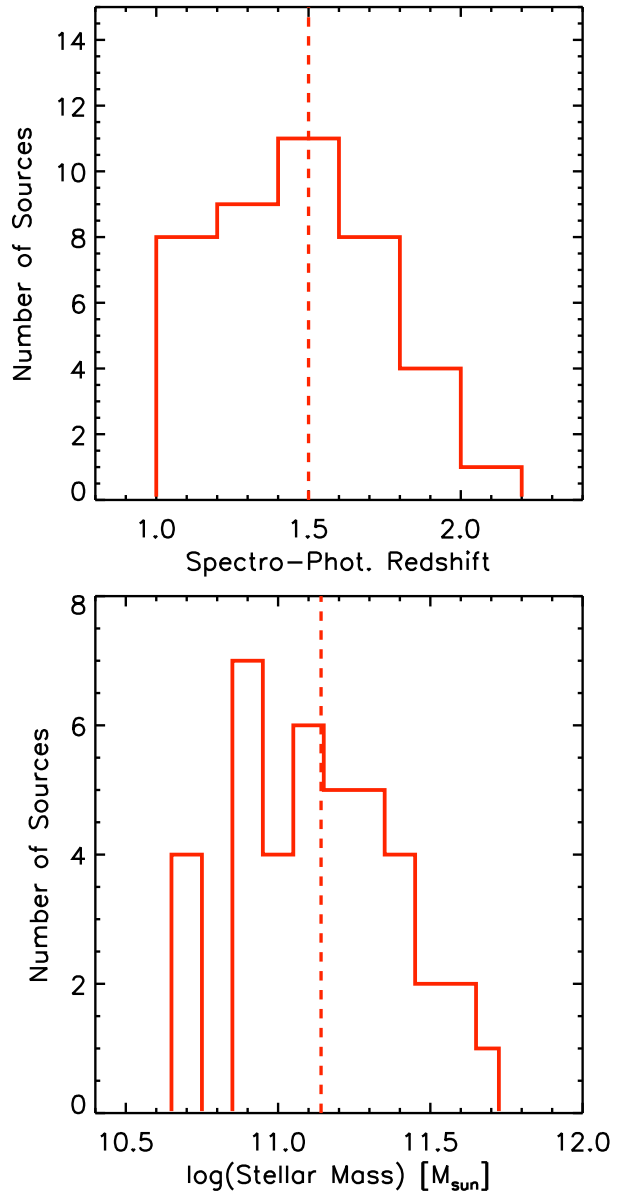}
\caption{For our sample of 41 quenched galaxies we show the histograms of redshift and stellar mass. Red dashed lines represent the median value for histogram.} 
   \label{fig:histoETG}
\end{figure}

\begin{figure}[]
   \centering
 \includegraphics[width=9.0cm]{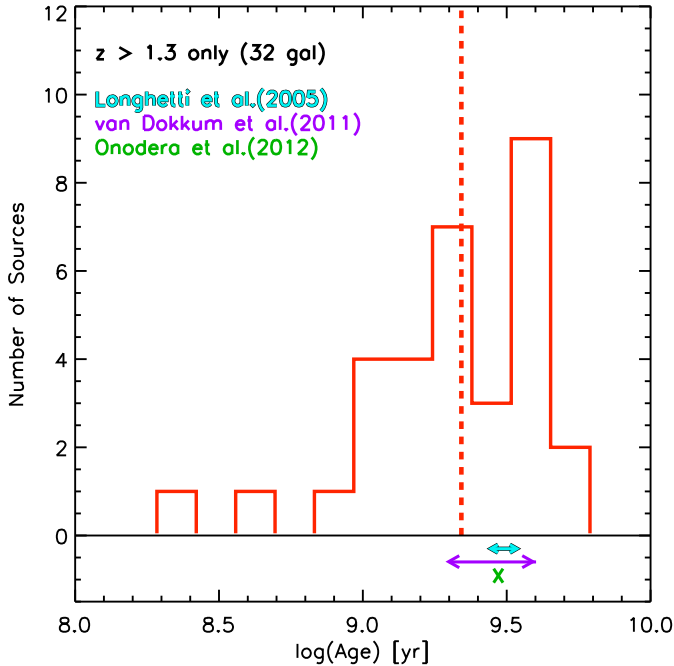}
 \caption{Histogram of luminosity-weighted age for 32 quenched galaxies at ${\it z} \geq 1.3$ (in red). Dashed line corresponds to the median of the distribution. We compare our median age for quenched galaxies with ages from early-type galaxies from literature: in turquoise, \citet{longhetti2005} mean age range (predictions from different models) of 10 ETGs at ${\it z} \sim 1.5$; in purple, \citet{vandokkum2011} age range of 15 low-$\rm H\alpha$-emission galaxies at $1 < {\it z} < 1.5$; in green the mean age of 18 ${\it z} \geq 1.4$ passively evolving galaxies from \citet{onodera2012}.}
   \label{fig:histAgeAv}
\end{figure}

\begin{figure*}[]
   \centering
  \includegraphics[width=18.0cm]{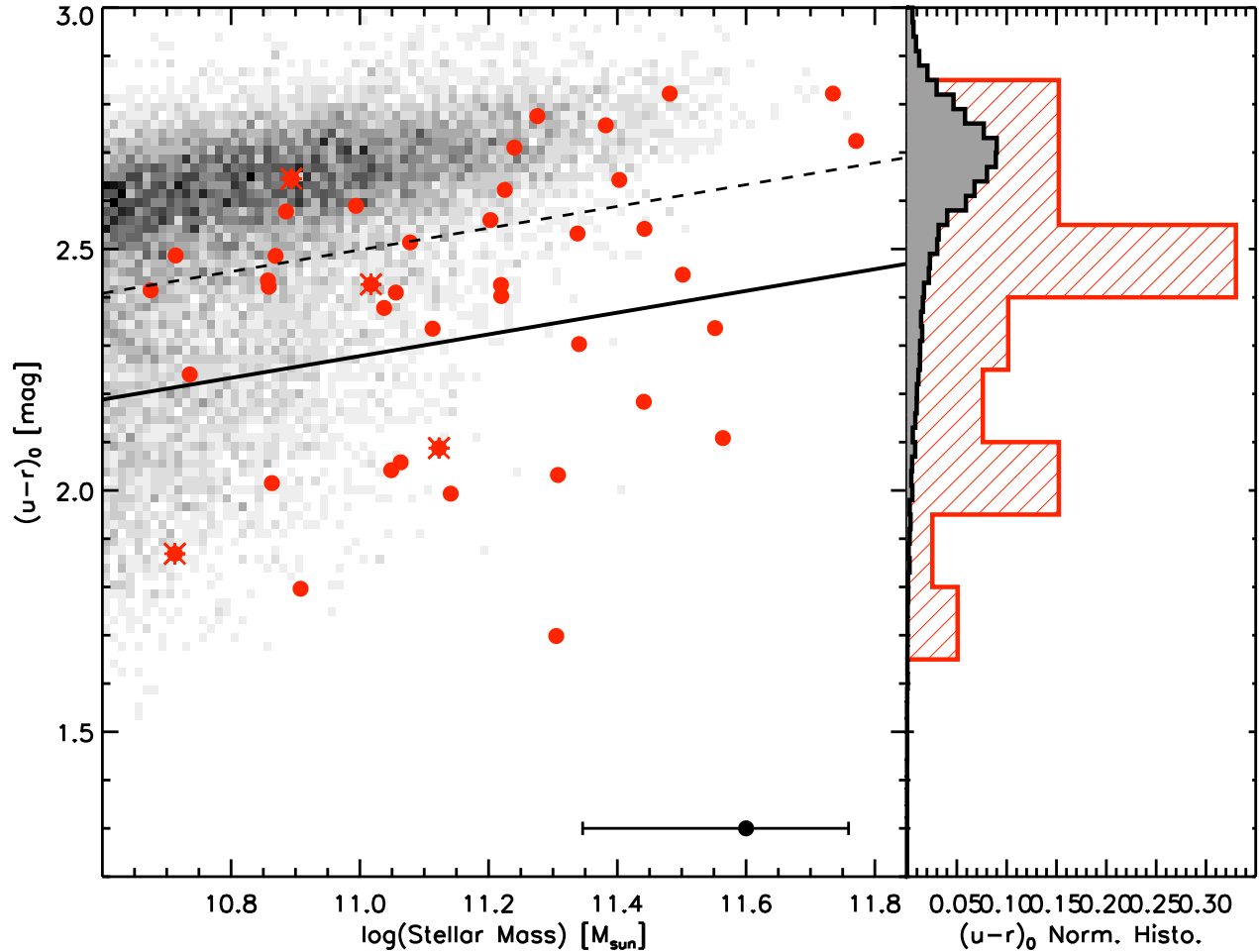}
 \caption{{\it Left}: In the rest-frame {\it u}$-${\it r} color versus stellar mass diagram we illustrate the large fraction of quenched (red) galaxies that have blue colors at these high stellar masses. Galaxies below the continuous tilted line are considered as ``blue''. This line represents a linear fit to the local SDSS ``red-sequence'' passively evolved to $z=1.5$ (dashed) plus an extra offset from the local red-sequence width (-0.42\,mag total offset). Asterisk symbols represent galaxies with emission lines. In gray, a density map with $\approx 11,000$ galaxies from SDSS-DR7 with ${\it z}\le 0.05$. In the bottom-right corner we show the median uncertainty in stellar mass for our galaxy sample. {\it Right}: We illustrate the ({\it u}$-${\it r})$_0$ color distributions for quenched galaxy samples. Overplotted a color histogram for SDSS galaxies with masses larger than our completeness mass limit. Each distribution is normalized to the total number of galaxies. Rest-frame colors were calculated from the best fit \citet{bc2003} model SED.}
  \label{fig:umr_mass}
\end{figure*}

\begin{figure}[]
   \centering
 \includegraphics[width=9.0cm]{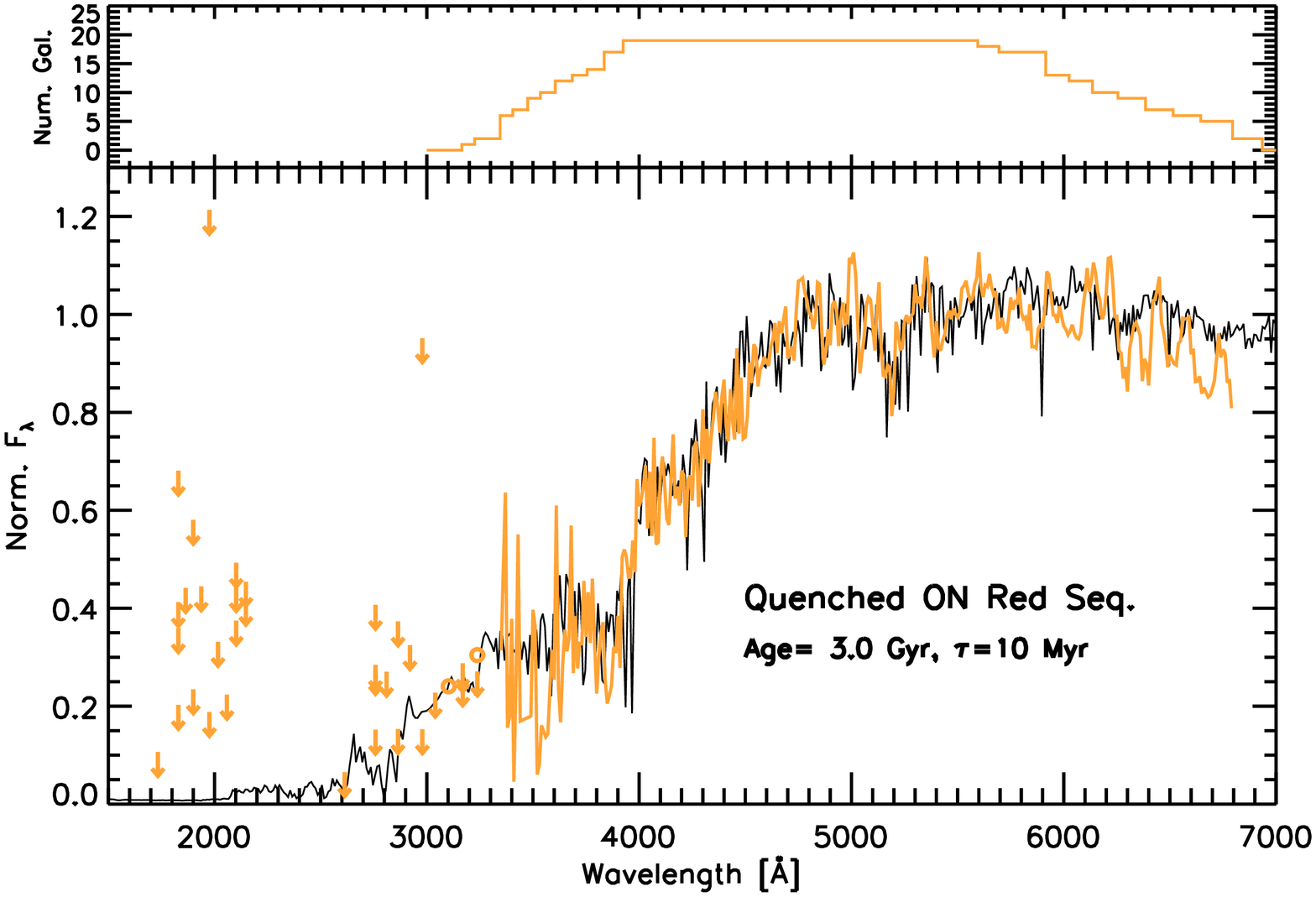}
  \includegraphics[width=9.0cm]{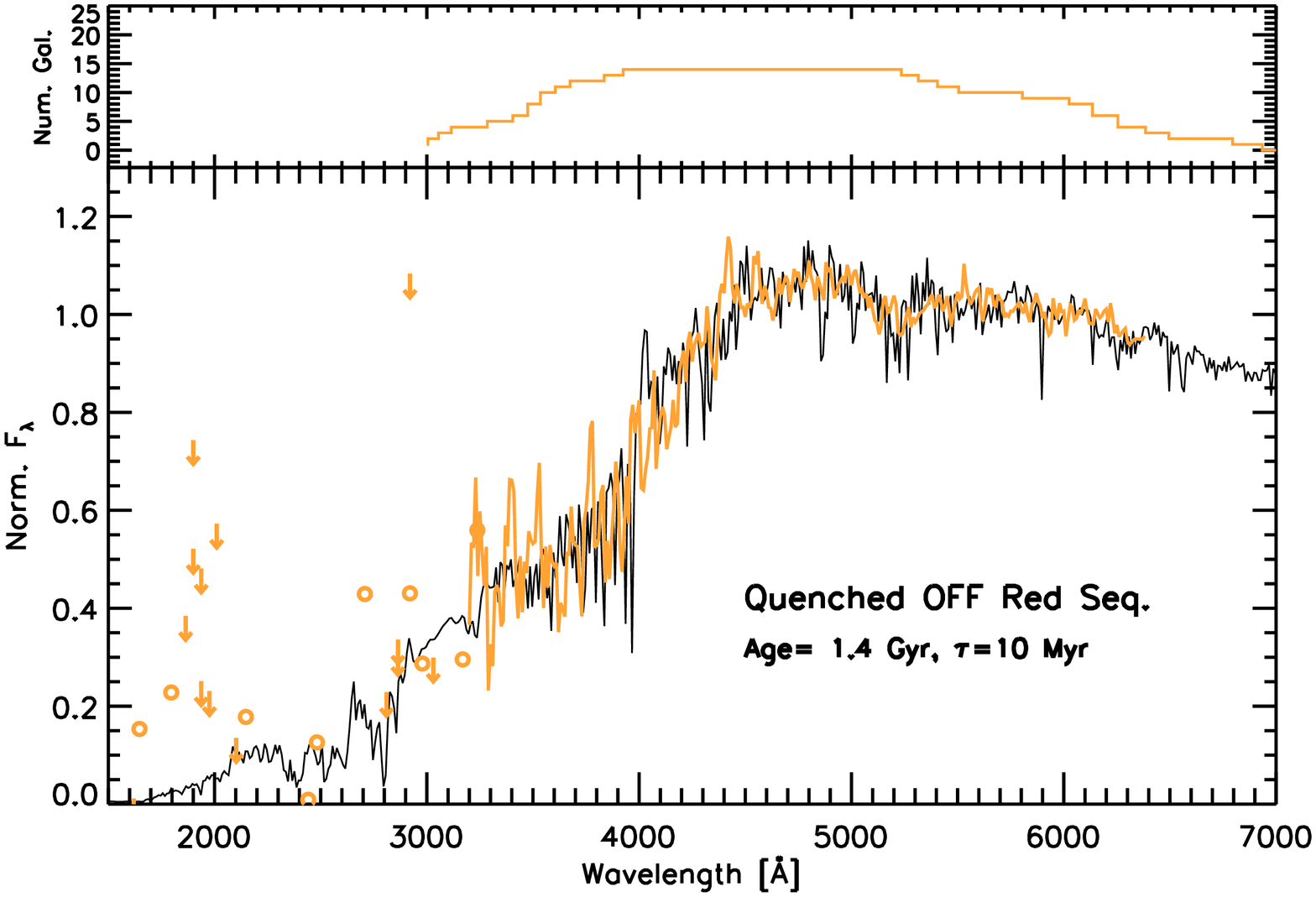}
 \caption{The stacked rest-frame spectra of ${\it z} \geq 1.3$ quenched galaxies emphasizes the differences between on-red-sequence sources (top, 19 sources) and off-red-sequence galaxies (bottom, 13 sources). Stacked spectra and UVIS data points for individual galaxies are shown in orange while best fit \citet{bc2003} models to the spectra are shown in black. The models are exponentially declining SFHs with $\rm \tau=10\,Myr$, solar metallicity, Salpeter IMF and ages of 3.0 and 1.4\,Gyr for on- and off-red-sequence galaxies, respectively. In the top of each panel we show the number of galaxies used in the coadded spectrum as function of wavelength. Flux densities normalized to median flux between 5500 and $\rm 6000\,\AA$.}
  \label{fig:coad}
\end{figure}

In the top panel of Figure~\ref{fig:histoETG} we show the distribution of redshifts for our quenched galaxy sample. Redshifts in our sample spread between 1.0 and $\sim 2.0$, with a median of 1.5. The sharp drop of the distribution for $z<1$ is a direct consequence of our selection against these galaxies. As described in Appendix~\ref{app:systerr} the redshift determination at $z<1$ is dominated by strong systematics, obviously affecting the derived SPPs.

We present the stellar mass distribution in the bottom panel of Figure\,\ref{fig:histoETG}. The stellar masses of quenched galaxies spread from $\rm log(M_{star}/M_{\odot}) = 10.65$ (our mass cut) to 11.70 with a median 11.15.

The SFHs of our quenched galaxies are consistent with short star-formation bursts, with all the galaxies having exponentially declining SFH with $\tau \le 100$ Myr.

In Figure\,\ref{fig:histAgeAv} we present the luminosity weighted age distribution for galaxies at ${\it z} \geq 1.3$ (where this parameter could be constrained, see Section\,\ref{sec:systerr} and Appendix\,\ref{app:systerr}). At these redshifts our sample consists of 32 quenched galaxies. We observe a wide distribution of ages between 1 and 4\,Gyr with a median of 2\,Gyr. We compare these results with data of quenched galaxies at similar redshifts from literature. The mean ages ($\rm \sim 3\,Gyr$) found by \citet{longhetti2005}, \citet{onodera2012} and \citet[][$2$-$\rm 4\,Gyr$ ]{vandokkum2011} are systematically above our median age, even though our age histogram shows a peak at 3-4\,Gyr consistent with these works. This may be due to different selections between our sample and those from literature. In particular, being selected as extremely red objects (EROs, $R-K' > 5$), the \citet{longhetti2005} sample is probably biased towards the oldest galaxies at these redshifts. The \citet{vandokkum2011} sample is based on WFC3 grism data as in our study (though they only have the G141 grism). However, their sample is mass-selected while we use mass {\it and} color to selected our quenched galaxies. Moreover, the reported ages in van Dokkum et al.\ are for galaxies significantly more massive than our sample ($\rm M_{star}>10^{11}M_{\odot}$).  Finally,  in order to maximize observing efficiency, the \citet{onodera2012} sample was constrained to galaxies inhabiting regions of high galaxy overdensity, where the star formation and quenching take place more quickly. Therefore, Onodera et al.\ sample would include mostly very old galaxies at the redshifts they were observed. Compared to $BzK$ $z>1$ galaxy selection (and alike), which separate SF and quenched galaxies \citep[e.g.][]{daddi2004,cameron2011}, our combination of spectral fitting and \jj$-$\hh\ color cut allows for blue (younger) quenched galaxies to be included in the sample (see Section below).

In summary, our wide age distribution cover the different age regimes previously reported in literature. Although our median age ($\rm 2\,Gyr$) is below previous reports, we have shown this does not imply an inconsistency between the different results but most likely it reflects different selection bias intrinsic to each galaxy sample.

\subsection{Mass versus rest-frame {\it u}$-${\it r} color} \label{discussion:mass_color}

In Figure\,\ref{fig:umr_mass} we show rest-frame {\it u}$-${\it r} color versus stellar mass for our sample of 41 quenched galaxies (in red, asterisks for galaxies with emission lines) together with local data from SDSS-DR7 \citep[grey region, ][]{abazajian2009}. SDSS colors for our galaxies were computed using the best-fit SED model and filter curves obtained from the SDSS webpage. When galaxies within our stellar mass range ($\rm log(M_{star}/M_{\odot}) \geq 10.65$) are considered, then the vast majority of SDSS galaxies belong to the red-sequence. The color distribution for these objects shows a narrow peak at ($u-r$)$_0\sim 2.7$. This is visible in the right panel of Figure~\ref{fig:umr_mass}, where we show the ({\it u}$-${\it r})$_0$ distribution of the local galaxy sample with $\rm log(M_{star}/M_{\odot}) \geq 10.65$ (grey histogram), together with the color distribution for our quenched galaxy sample. Our massive galaxies show comparatively a much broader color distribution than massive galaxies in the local Universe, with ($u-r$)$_0$ colors as blue as 1.7 \citep[consistent with recent results, see e.g.,][]{vandokkum2011,newman2012}.  We note that, because of the color preselection applied to our galaxy sample we are excluding even bluer, higher SFR galaxies. Therefore the number of blue massive galaxies we obtain can be considered as a lower limit, implying an even large spread in ({\it u}$-${\it r})$_0$ colors at these redshifts compared to the local massive galaxy sample.

\subsection{Young and old quenched galaxies} \label{discussion:yando}

In Figure\,\ref{fig:umr_mass} we find that our quenched-galaxy color distribution peaks at ({\it u}$-${\it r})$_0= 2.5$, approximately 0.2 magnitudes bluer compared to the peak of the color distribution of local red-sequence galaxies. This color difference matches the passive evolution of a short burst of star formation expected between ${\it z} \sim 1.5$ (the median redshift of our galaxy sample) and ${\it z} \sim 0$. 

About 32\% of our sample, however, has colors substantially bluer ($\geq 0.4$ magnitudes) than the local red-sequence, overlapping with the colors of local blue-cloud galaxies at masses below M$\rm _{star} \sim 3\times10^{9}$ M$_{\odot}$.  In Figure~\ref{fig:umr_mass} we show the best-fit red-sequence to the ${\it z}=0$ SDSS data passively evolved to ${\it z} \sim 1.5$ (0.2\,mag offset, dashed line) and with an additional offset so galaxies fall within the scatter of the red-sequence (0.22\,mag, continuous line).  The SDSS red-sequence was computed by fitting a linear relation to the SDSS log(M$_{\rm star}$) and ({\it u}$-${\it r})$_0$, while the width was taken as the 2$\sigma$ dispersion with respect to this fit. We use this line to split our sample in quenched galaxies {\it on} and {\it off} the ${\it z} \sim 1.5$ red-sequence. Approximately $1/3$ of the sample is {\it off} the red-sequence.

At the beginning of this Section we describe how our \jj$-$\hh\ selection method produces a quenched galaxy sample with lower median age than previous works in literature. We also describe how the selection effects of each of the cited works systematically bias their samples towards older ages. Driven by this result we further investigate on the ages of our two sub-samples: quenched galaxies on and off the red-sequence. Limiting the sample to quenched galaxies at ${\it z} \geq 1.3$ (where the age-sensitive 4000\AA/Balmer break is covered by the spectroscopy), we find a significant age difference between the stellar populations of galaxies on and off the red-sequence. The red-sequence quenched galaxies have a median age of 3.1\,Gyr, while quenched galaxies off the red-sequence have a median age of only  1.5\,Gyr.

We further investigate this age difference by stacking the spectra of the quenched galaxies on and off the red-sequence in order to increase our S/N. Figure\,\ref{fig:coad} shows the coadded spectra of quenched red-sequence (top, 19 sources, orange) and off-red-sequence galaxies (bottom, 13 galaxies, orange), together with their best fit \citet{bc2003} stellar population models (black curves). The stack data were fitted using our full model library, where only metallicity was fixed to solar. The best fit models are exponentially declining SFHs with $\rm \tau=10\,Myr$, solar metallicity, Salpeter IMF and ages of 3.0 and 1.4\,Gyr for on- and off-red-sequence galaxies, respectively. As we see these models provide a good representation of the stacked data. Also they confirming our results from the two galaxy sub-samples based on individual-galaxy analysis. These results are consistent with recent findings of \citet{whitaker2013} in coadded spectra, even though their ages for blue ($\rm 0.9\,Gyr$) and red quenched galaxies ($\rm 1.6\,Gyr$) are younger than ours.

\subsection{``Rejuvenation'' of old quenched galaxies through a secondary star-burst?} \label{discussion:rejuv}

The \citet{bc2003} models we considered in Section~\ref{discussion:yando} assume a SFH with a single massive star- formation events that passively declines with time. In this scenario, a possible interpretation of the distinction between the measured ages of galaxies on and off the red-sequence arises from the progress of the sample towards the red-sequence observed at different epochs of evolution for individual galaxies.

Even though these models fit well the coadded spectra, these are not the only possible SFHs for these galaxies. The presence of small secondary star-bursts (SB) after quenching of the main stellar component have been observed at low and high redshifts \citep[e.g.,][and references therein]{kaviraj2013}. In this section, we consider the possibility that our quenched galaxies lie {\it off} the red-sequence due a relatively recent minor burst of star formation which rejuvenates the galaxies' SEDs, making them appear relatively younger due to the production of luminous, high-mass O/B/A/ stars. 

To test this scenario we constructed a new set of model SFHs, which are defined by a primary star-formation event (fixed at very high-$z$) followed by a smaller, secondary SB. Both, the main stellar component and the SB are modeled with exponentially declining SFRs, with solar metallicity, Salpeter IMF and no extinction. The main component produces a total stellar mass of $\rm 10^{11}$~M$_{\odot}$ with $\rm \tau=10\,Myr$. The age of this main component before the SB ranges between 0.5-4.25\,Gyr in steps of 0.25\,Gyr (16 model variants). The secondary SB has intensities of 1, 5, 10 and 50\% the total galaxy mass (4 model variants) with $\rm \tau=100\,Myr$. We observe the resulting SEDs at the time of the SB peak and after 50, 100, 250 and 500\,Myr of that event (5 model variants). The combinations of these possible parameters make a library of 320 models to be fitted to the coadded data.

\begin{figure}[]
   \centering
  \includegraphics[width=9.0cm]{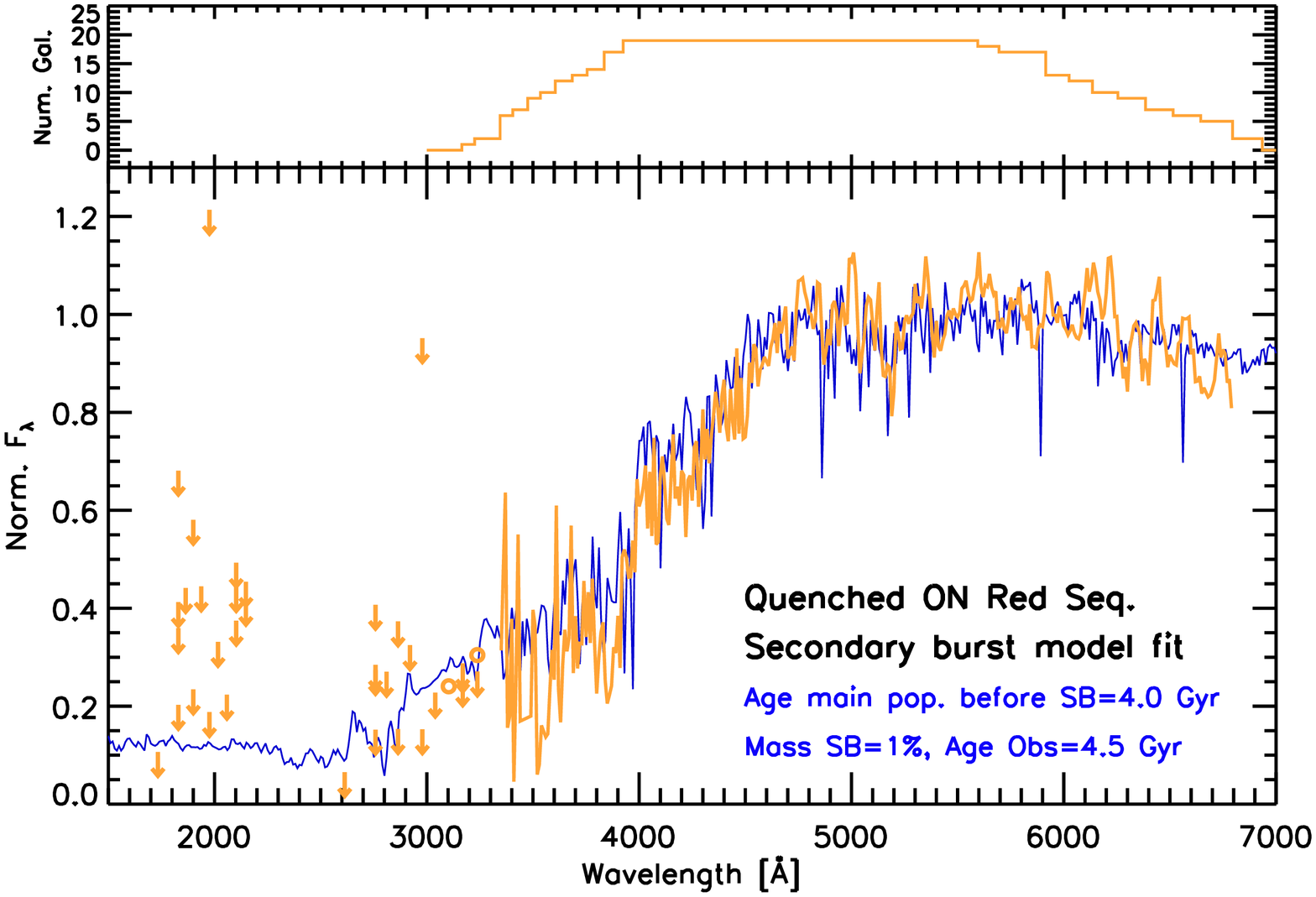}
  \includegraphics[width=9.0cm]{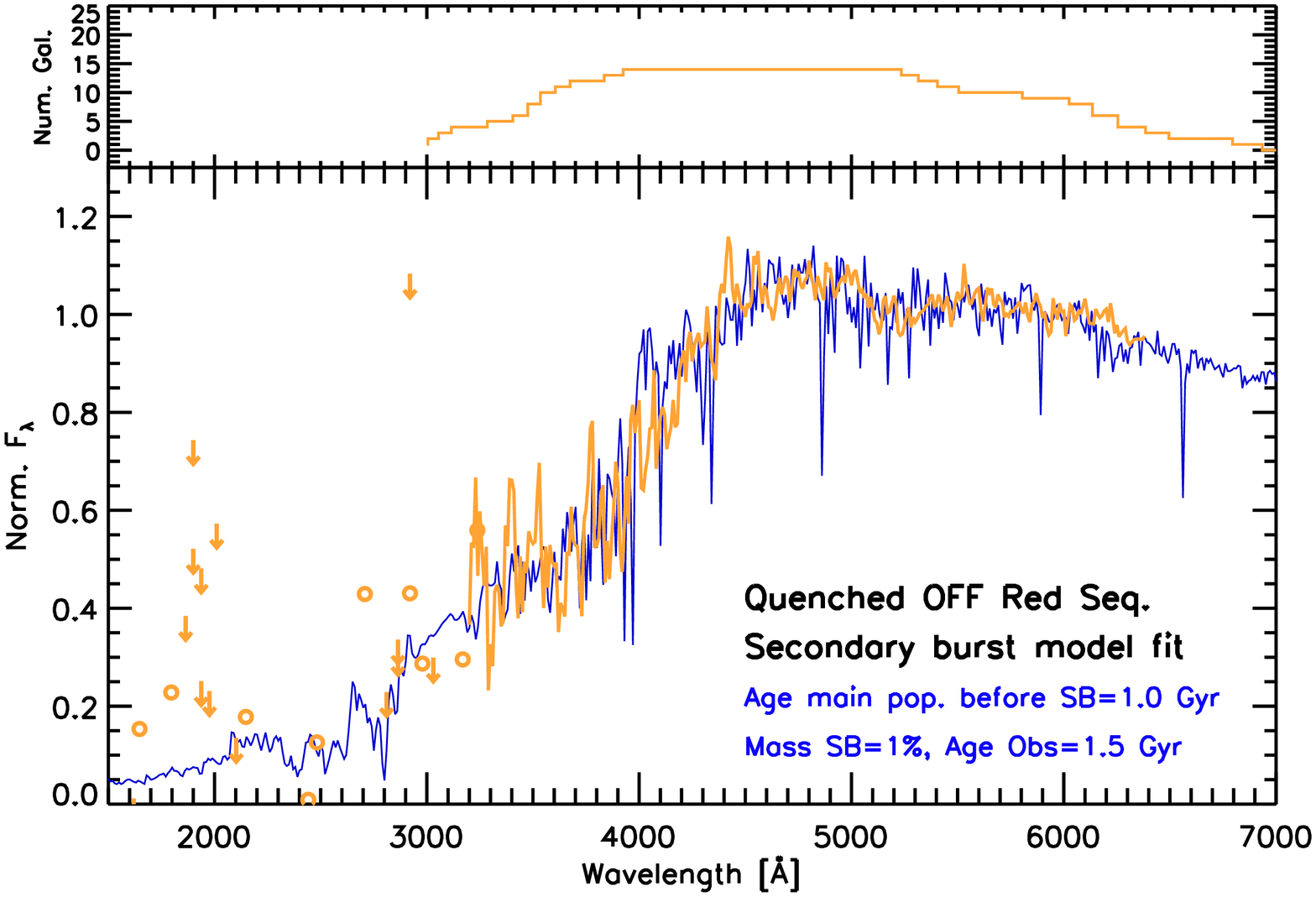}
  \includegraphics[width=9.0cm]{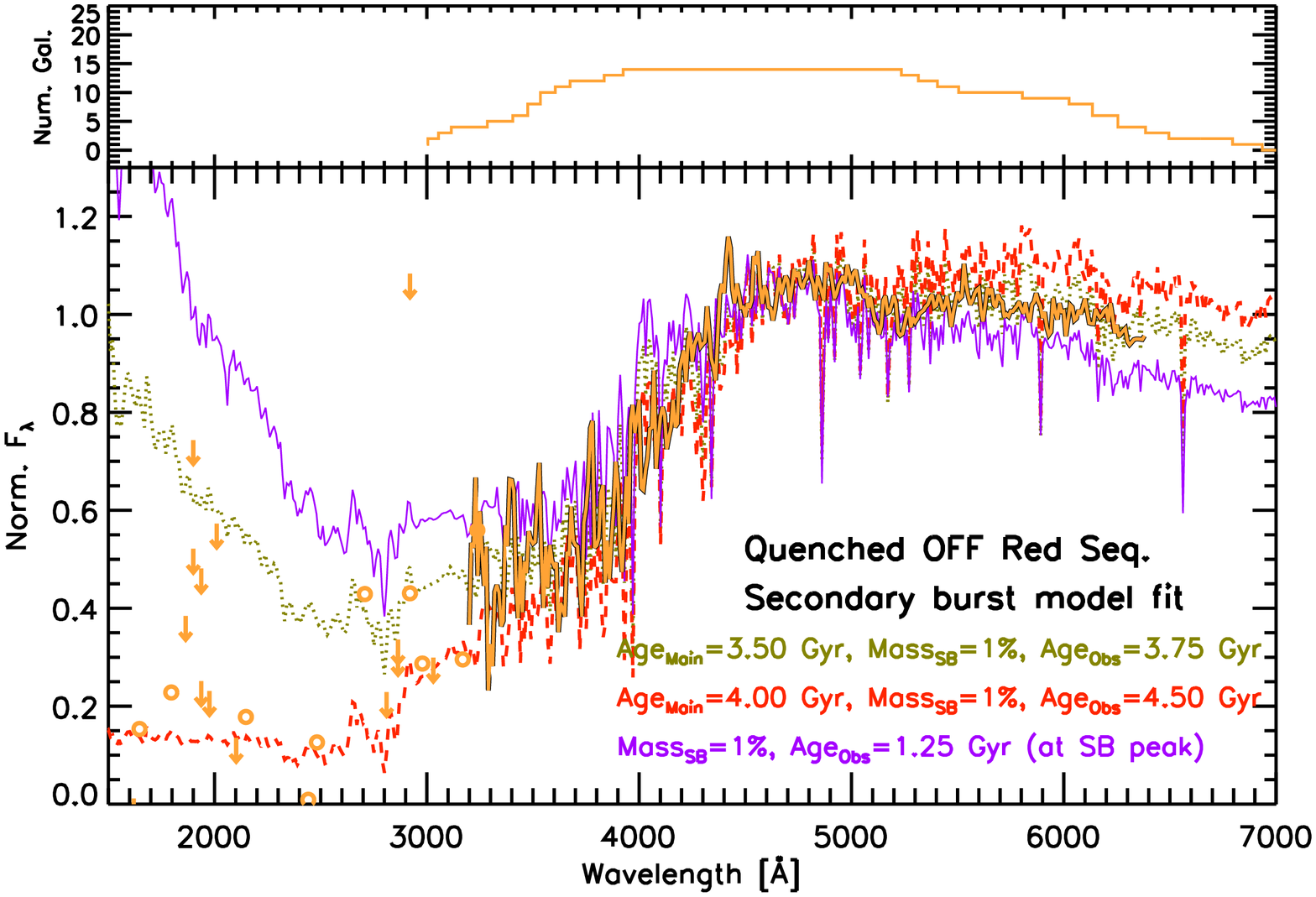}
 \caption{Same stacked data as in Figure\,\ref{fig:coad} (orange). {\it Top panel:} coadded spectra and individual-galaxy photometry of quenched galaxies on the red-sequence together with the best model fit with two stellar components (main+SB, in blue).  {\it Middle panel:} Our two-stellar-component models can not reproduce a "rejuvenation'' SFH (old main component + young SB) consistent with the data. We show the coadded data of quenched galaxies off the red-sequence together with the best model fit with two stellar components (both components young, in blue).  {\it Bottom panel:} Off the red-sequence galaxy data together with three models: in purple, best model fit among those observed at the SB peak. In green, best model among those with ages of observation $\rm \geq 3 Gyr$. In red,  best model found for galaxies on the red-sequence (top panel), with the proper scaling factor to better fit the off red-sequence data.}
  \label{fig:coadSB}
\end{figure}

In Figure~\ref{fig:coadSB} we present the best model fits for our coadded spectra of quenched galaxies on and off the red-sequence (orange).

Just as a point of reference, in the top panel of Figure~\ref{fig:coadSB} we see (in blue) the best model fit for quenched galaxies {\it on} the red-sequence that includes a secondary SB. This model has an age of 4.5\,Gyr. The main stellar component is 4.0 Gyr old before the SB, which burst mass is 1\%. The coarse shape of the coadded spectra is fitted by this model, showing however, a larger $\chi^2$ value than the single-burst best model (top panel Figure~\ref{fig:coad}). The predicted 4.5\,Gyr is older than the 3\,Gyr from the single-burst model. This is not surprising as the galaxy must passively evolve after the SB to recover the originally red SED.
 
In the central panel of Figure~\ref{fig:coadSB} we present the best model fit to the coadded spectra of quenched galaxies {\it off} the red-sequence. This model has an age of 1.5\,Gyr. The main stellar-component age before the SB is 1.0 Gyr, with a burst mass of 1\%. The predicted age of 1.5\,Gyr is consistent with the single-burst model prediction for these galaxies. We highlight that the best model for the quenched off red-sequence galaxy data is not a combination of an old, passive stellar population with a younger SB. The fitting process demands both stellar populations to be similarly young, resembling a single-component young burst, i.e., the ``rejuvenation'' scenario is not supported by this analysis.

Given the above result, we proceed by being more selective on choosing models that resemble a ``rejuvenation'' SFH, and test if they fit the data of off red-sequence galaxies. In the bottom panel of Figure~\ref{fig:coadSB} we show three examples. In purple we plot the best model fit to the coadded spectrum for all models observed at the peak of the SB. The main stellar component is, again, young (1.5\,Gyr). More important, even the weakest SB we are testing (1\%) can not reproduce the data and clearly under-predicts the flux in the red extreme of the coadded spectrum. Also, the blue flux is over-predicted as it can be seen by comparing the best fit models to the individual-galaxy UVIS data points (not included in fit). Only one upper-limit among 26 data points is consistent with this blue SED model.

In addition, we also considered only those models with old ($\rm age \geq 3 Gyr$) primary stellar component. The resulting best-fit model is shown in green in Figure~\ref{fig:coadSB}. Similarly to the previous library, also this best-fit model over-predicts the flux at these blue wavelengths, and it is consistent with only four out of 26 observed UVIS data points. Finally, we overplot in red the best fit model found for galaxies {\it on} the red-sequence, with the proper minimization-scaling factor to fit the {\it off} red-sequence data. This model is reasonably compatible with the individual-galaxy UVIS data (as before, not included in fit), but it clearly over-predicts the flux in the red extreme of the coadded spectrum.

In summary, after testing a library of models that include a secondary SB, we have found that they can not reproduce the ``rejuvenation'' scenario for most of our {\it off} red-sequence galaxies. The better fits are produced with overall young stellar populations (for both, main component and SB), resembling a single-young stellar component. The use of $\rm \geq 3 Gyr$ main stellar components systematically over/under predict the data flux at different wavelengths. We remind the reader, however, we have not consider extinction in the model library used in this section. As we have seen before, extinction is a poorly constraint parameter that also introduces strong degeneracies with other SPPs. Therefore, the results shown in this section must be taken with caution, and restricted to the model library we use here.

\subsection{Red fraction of massive galaxies at $z \sim 1.5$} 
We also investigate the fraction of quenched galaxies which have already settled on the red-sequence ($\rm f_{red}$) and those which have not, at the average redshift of our sample ($z\sim 1.5$). 
Using the red-sequence computed above, we found that  28/41 galaxies fall within 2$\sigma$ of the red-sequence. These galaxies comprise 68\% of our whole sample. Since all blue SF galaxies are excluded from our selection, we infer that  less than $68\%$ of all galaxies at ${\it z} \sim 1.5$ with $\rm log(M_{star}/M_{\odot}) \geq 10.65$ fall on the red-sequence.
We stress that this is an upper limit. While our \jj$-$\hh\ color pre-selection allows us to select all of the red galaxies at $z \gtrsim 1.0$, it systematically excludes blue galaxies at the same redshifts. 

We can improve the upper limit on $\rm f_{red}$ by including emission-line-selected sources in the same apparent magnitude, redshift and mass ranges of our sample.  For this, we use the \citet{dominguez2012} sample of emission-line galaxies from 17 WISP fields (all fields in common with our sample). These galaxies are selected only by the presence of the \ha\ emission line.  Dominguez et al.\ limit their search to $z=1.5$, thus probing a smaller volume compared to our study. As a result the value of  $\rm f_{red}$ provided  below has still to be considered as an upper limit. We applied the same \hh-band magnitude, mass and redshift selection criteria used in our sample galaxies, to identify sources within our parameter ranges. This selection results in 15 galaxies which are not in common with our galaxy sample. Stellar masses from Dominguez et al.\ were scaled by a factor of 1.7 to account for the different IMFs used (Chabrier in their case, Salpeter in ours). By scaling the number of emission-line sources to our number of fields, we derive an expected number of 24 blue massive emission-line galaxies.  This implies that, at $z\sim 1.5$, $\rm f_{red} < 43\%$. 

Recently, \citet{peng2010} predicted the evolution of the red fraction as a function of galaxy mass and redshift, based uniquely on empirically-motivated relations. Peng et al.\ predict that at ${\it z} \sim 1.5$, at the median mass of our galaxy sample, between 70 to 90\% of the galaxies should have colors consistent with being on the red-sequence. These fractions were derived from their predictions for galaxies in low-density environments\footnote{The red-galaxy fraction for galaxies in high-density environments is predicted to be larger.}. 

We also note that, even though the typical mass of our sample is well within what Peng et al.\  identify as the ``mass quenching'' regime, (i.e., quenching depends only on intrinsic properties of the galaxy and {\it not} by the local environment it inhabits) our color-mass relation in Figure\,\ref{fig:umr_mass} does not show such a mass dependency. In our data, whether a quenched galaxy resides on or off the red-sequence appears to be independent of the galaxy's mass. On the contrary, a mass-dependent quenching mechanism would demand a rising left-to-right gradient in the color-mass diagram, such as the most massive galaxies mostly populate the red-sequence while less massive systems should be bluer. This is not observed in our galaxy sample, despite covering one order of magnitude in mass. Therefore, it seems reasonable to conclude that both internal (e.g., mass quenching) and external (e.g., environment) mechanisms are at play in quenching the star formation in some of these massive galaxies.

We speculate that the young massive galaxies with no on-going star formation observed at ${\it z} \sim 1.5$ may be in transition between a phase of vigorous star-formation at ${\it z} >2$ and the $z\sim 1.5$ red-sequence.  The large masses and short SF-timescales that characterize these galaxies imply very high past SFRs (of the order of $\sim 1,000$\,M$_{\odot}$ yr$^{-1}$). The survey volume covered in the $1.3<{\it z} <1.7$ redshift range is a few $10^8$\,Mpc$^{3}$, implying a volume density for our quenched galaxies of $\sim 10^{-7}$\,Mpc$^{-3}$. Even considering that we had to remove from our sample about half of the galaxies because of spectral contamination, this volume density is still comfortably below the observed space densities of some of the likely progenitors of these sources, i.e., ultra luminous IR galaxies at $z \sim 2.5$ \citep[6$\times 10^{-6}$\,Mpc$^{-3}$][]{chapman2005}.\\\\

 Finally, we would like to highlight the results presented in this paper do not change strongly because of a misclassification of emission-line galaxies as ``quenched''. If we remove the 4 quenched galaxies with emission from the sample we find the following results. First, for quenched galaxies on and off the red-sequence the median ages ($\rm 3.1\,Gyr$ and $\rm 1.3\,Gyr$) and single-component ages from coadded spectra ($\rm 3.0\,Gyr$ and $\rm 1.3\,Gyr$) are virtually unchanged with respect to our complete sample estimations. Second, we find that 72\% (68\% previously) of the sample galaxies are on the red-sequence. Also the $\rm f_{red} < 43\%$ does not change with respect the upper limit found previously. In consequence, we can be confident that neither the results not the global picture presented in this work are affected by SF galaxies misclassified as quenched systems.

\section{Conclusions}
We have used the WISP survey to identify a sample of massive ($\rm log(M_{star}/M_{\odot}) \geq 10.65$) galaxies at redshift  $\rm 1.0 \leq {\it z} \lesssim 2.0$. The sample was selected to have \jj$-$\hh$\geq 0.6$ and \hh$\geq 23$, in 27 independent fields, overcoming the effect of cosmic variance, which typically plagues massive galaxy samples at these redshifts.  Our color selection implies that the final sample is biased against star-forming galaxies.
We derived stellar population parameters by fitting stellar population models to the combined broad band photometry (HST/WFC3-UVIS, HST/WFC3-IR, {\it Spitzer}/IRAC) and IR spectra  ($0.9\,\mu$m\,$<\lambda<1.6\,\mu$m). We have shown that the availability of rest-frame optical spectra covering the 4000\AA\ /Balmer breaks dramatically  improves the reliability and accuracy of the derived stellar population parameters. 

Our results are based on a color- {\it and} mass-selected sample 41 quenched galaxies. In agreement with other studies, we find that, at $\langle z \rangle = 1.5$,  the mass range above $\rm log(M_{star}/M_{\odot}) = 10.65$ is populated by galaxies with a wide range of stellar population properties.  We find that quenched galaxies are well fitted with exponentially decreasing SFHs, and short star-formation timescales ($\rm \tau \le 100\,Myr$). They also show a wide distribution in stellar ages, between 1-4 Gyr.

We find that quenched galaxies are far from being an homogeneous population. In the $(u-r)_0$-versus-mass space, quenched galaxies have a large spread in rest-frame color at a given mass. Most quenched galaxies populate the ${\it z}\sim 1.5$ ``red-sequence'', although 32\% of them have substantially bluer colors. We find that quenched galaxies {\it on} the red-sequence have older median ages (3.1\,Gyr) than the quenched galaxies {\it off} the red-sequence (1.5\,Gyr). The average ages of the two subsamples (on and off the red-sequence) are confirmed by the analysis of their stacked spectra. Furthermore, we also demonstrated that a ``rejuvenated'' SED cannot reproduce the observed stacked spectra.  

We derive the upper limit on the fraction of galaxies on the red-sequence at $z\sim 1.5$ to be $\rm f_{red} < 43\%$, in disagreement with empirical model predictions from Peng et al.\ (2010, $\rm f_{red}=$70-90\%). This mismatch can partially be due to the Peng et al.’ assumption of an instantaneous quenching mechanism. However, the homogeneous spread in mass of our quenched galaxies on and off the red-sequence suggests that more than one mechanism is responsible of the quenching at these stellar masses (i.e., internal, galaxy-mass dependant vs.\ external, environmental triggers).

 We speculate that the young massive galaxies with no on-going star formation are in a transition phase between vigorous star formation at $z>2$ and the $z \sim 1.5$ red-sequence. According to their estimated ages, the time required for quenched galaxies {\it off } the red-sequence to join their counterparts {\it on} the $z\sim 1.5$ red-sequence is of the order of $\rm \sim 1\,Gyr$. 

The open question remains concerning what {\it mechanisms} halts the star formation. Having been quenched more recently, the galaxies {\it off} the red-sequence with no on-going star formation will be the ideal laboratory to further investigate this process. Once the WISP survey is completed, we will be able to study this and other galaxy populations in greater detail, with a much larger number statistics.

\acknowledgments
The WISP survey is supported by grants
HST-GO-12283 and HST-GO-12568 awarded by the Space Telescope Science Institute.
CLM acknowledges support from NSF grant \# AST-1109288.

\appendix

\section{The UVIS Charge Transfer Efficiency correction} \label{app:cte}
WFC3/UVIS CCDs experience a degradation of their Charge Transfer
Efficiency (CTE) over time, introduced by their exposure to energetic
radiation. The CTE degradation and fractional associated losses worsen
for low sky background, faint fluxes, and distance from the readout
amplifiers. These effects are discussed in \citet{noeske2011} for
observations obtained between October 2009 and October 2011.
Correction for CTE losses is currently not implemented in the
reduction packages for the WFC3-UVIS data, so we corrected the
measured fluxes as follows.

\cite{noeske2011} provides linear fits of the CTE losses as a
function of background electron counts, position on the detector, and
source flux. We interpolated the linear fits to compute, for
each galaxy, a unique correction as a function of the source counts
and position on the CCD. For sources fainter than 500 e$^-$ the
correction was obtained by extrapolating the solution from
higher counts;
while no correction was applied for sources brighter than 16,000
electrons. As the source detection for our photometric catalogue was
performed in the deeper (and brighter) $J_{F110W}$ images, many of the
WFC3/UVIS fluxes are below the extrapolation limit of 500 e$^-$ ($\rm
\approx 45 \%$ of the catalogue sources). Typically, these sources
have median magnitudes between 25.0-25.5 in the different UVIS bands
used in this study. Most of these sources ($\approx 95 \%$), however,
are also below our $\rm 3 \sigma$ detection limit, so upper limits for
their magnitudes (independent of the CTE correction) are used instead 
(see Section\,\ref{sec:photcat} for details). Therefore, in practice most of the sources with $\rm 3 \sigma$
are above the extrapolation limit for the CTE correction.

\begin{figure}[!t]
   \centering
  \includegraphics[width=8.5cm]{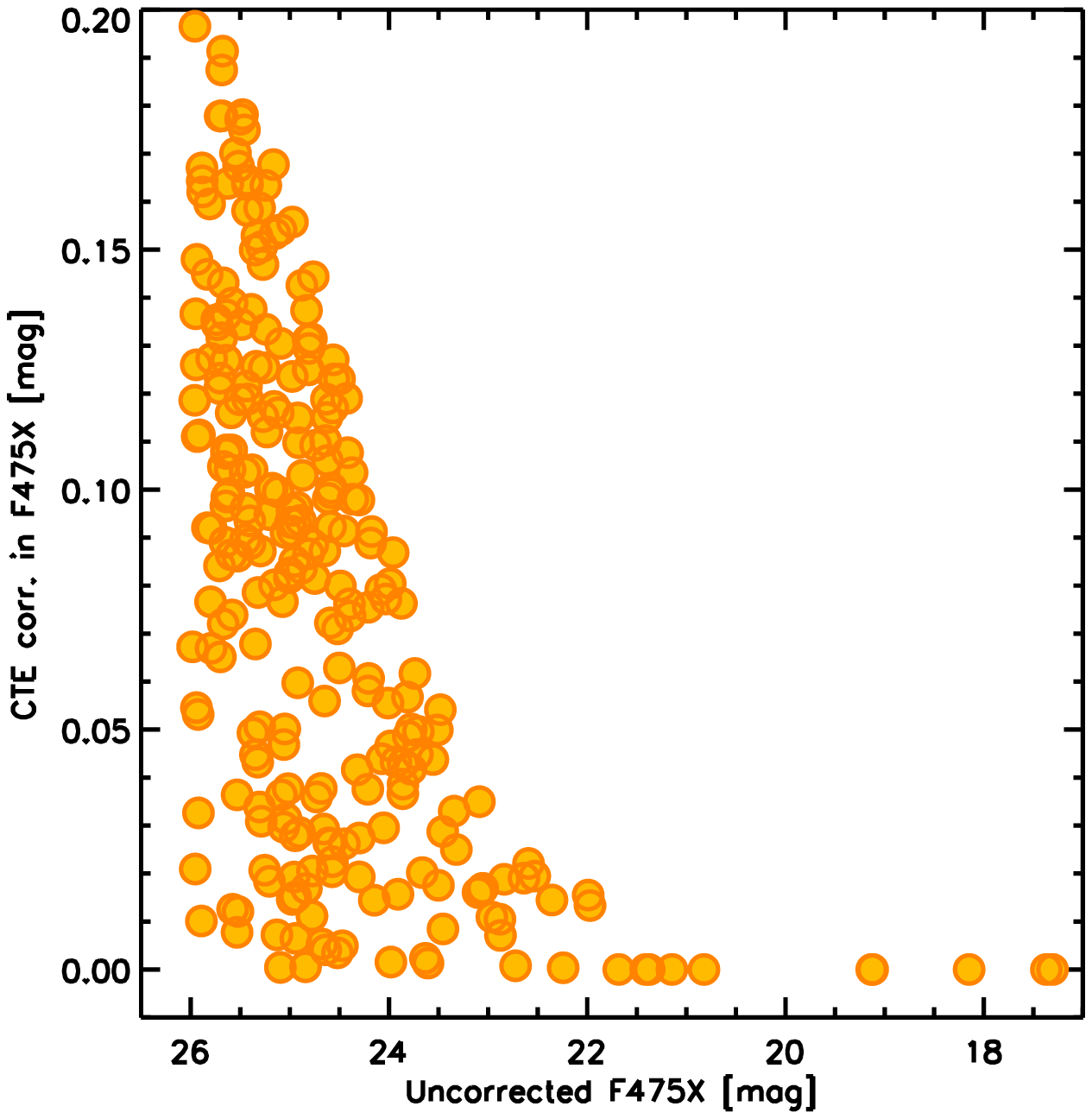}
   \includegraphics[width=8.5cm]{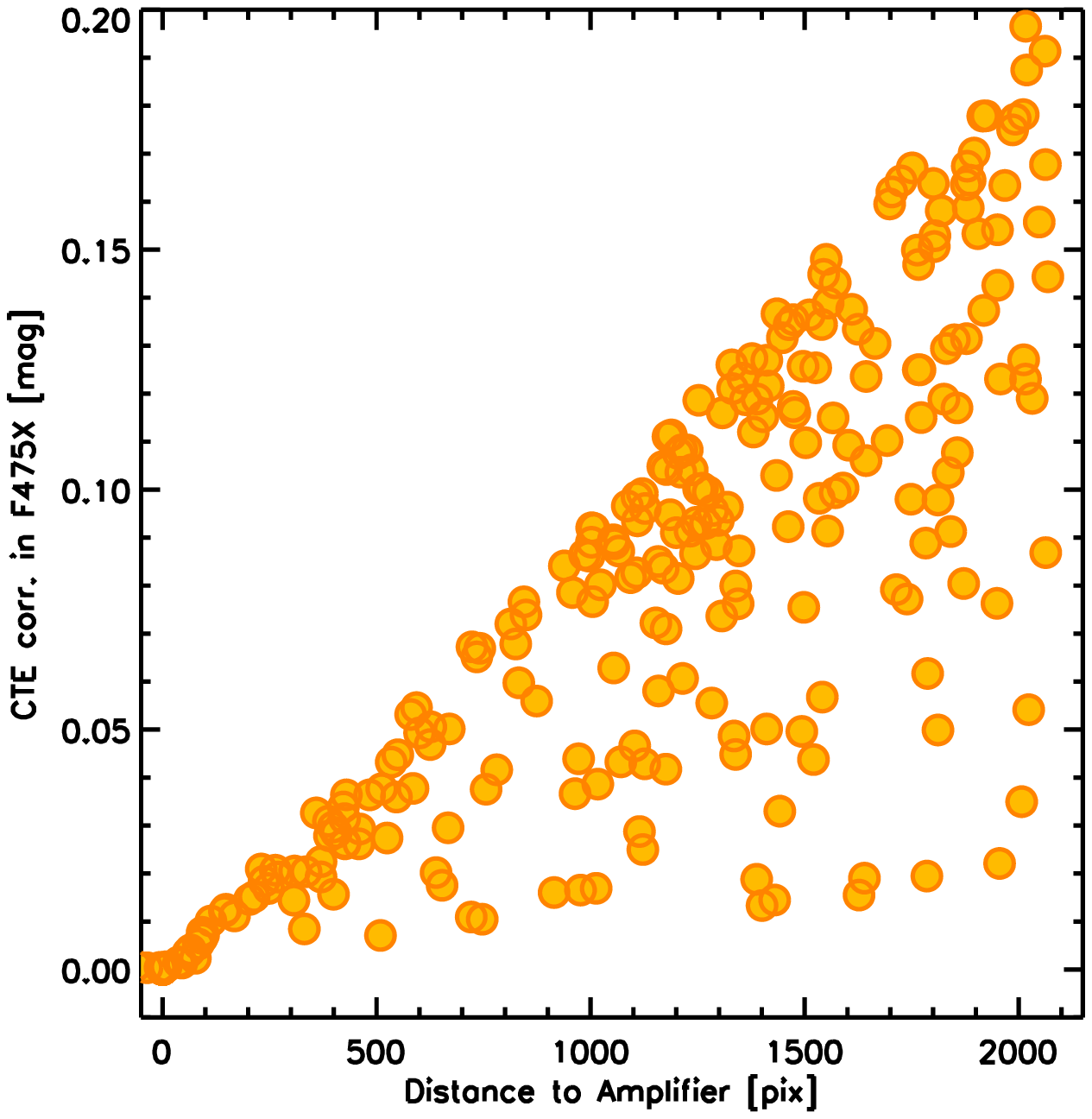}
  \caption{Example of Charge Transfer Efficiency (CTE) correction in field
    Par83. {\it Left:} In the CTE correction versus F475X magnitude
    (uncorrected) plot we illustrate that for most of our sample galaxies the corrections are below 0.1\,mag ($<5\%$ their total fluxes); {\it Right:}
    In the CTE correction versus distance to the amplifiers [pixel] plot we see that the correction is usually larger as the distance to the amplifier increases.}
  \label{fig:cte_corr}
\end{figure}

In Figure~\ref{fig:cte_corr}, we show an example of the CTE
correction in one of our fields versus the original source magnitude
(left) and versus Y-axis distance with respect the amplifiers
(right). Most of our final galaxy sample (see
Section~\ref{sec:selectETG}) have corrections below 0.1\,mag, typically
corresponding to $<5\%$ in their total fluxes.

\section{Spectra extraction} \label{app:extraction}
\begin{figure*}[!t]
   \centering
  \includegraphics[scale=0.45]{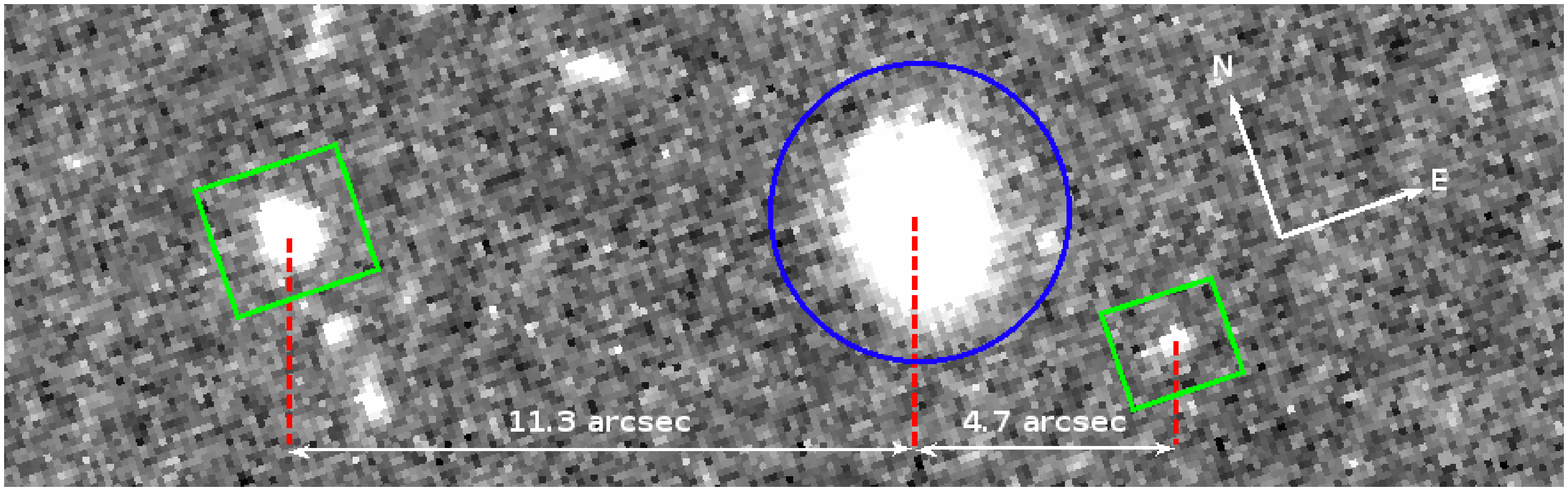}
   \includegraphics[scale=0.35]{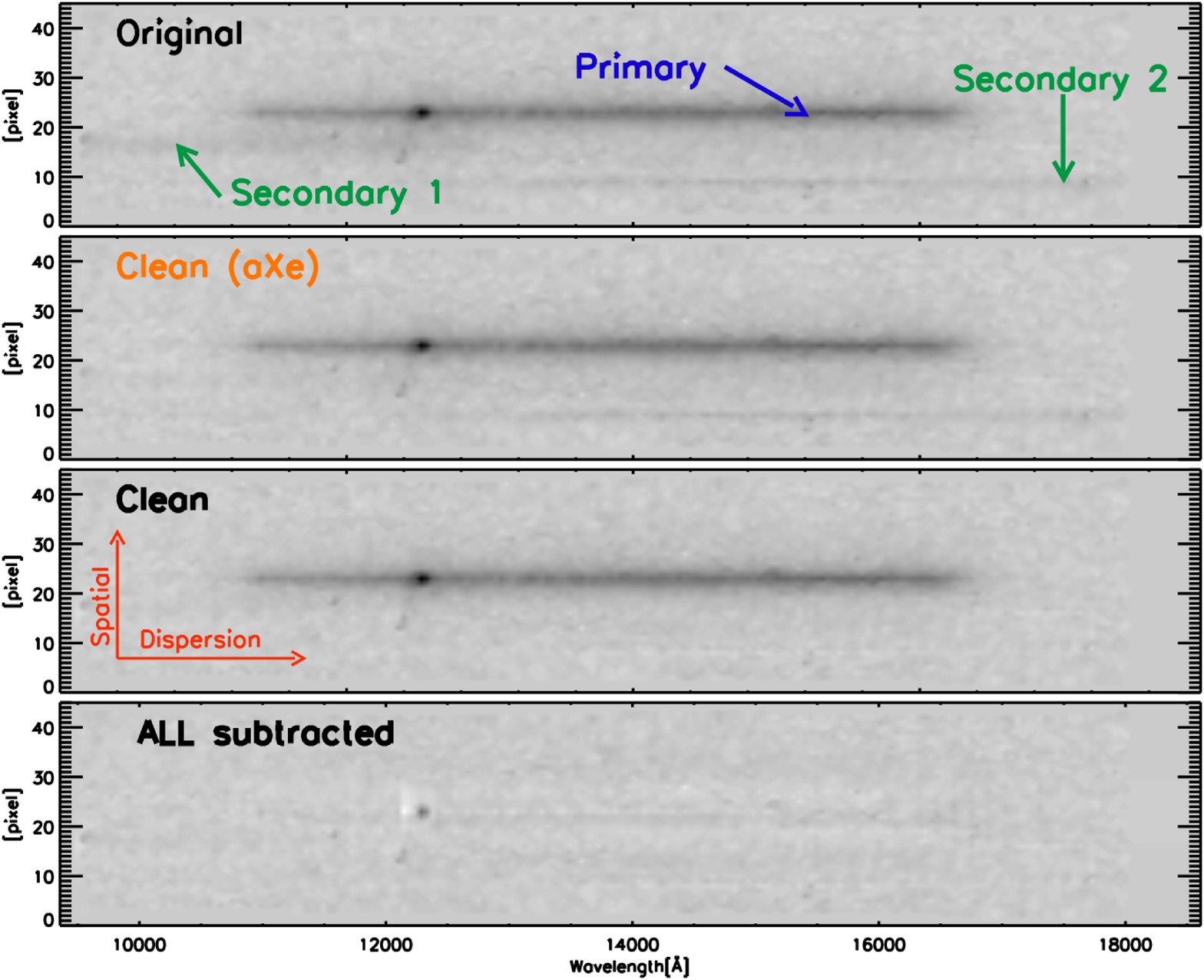}
\caption{ Example of 2D spectral stamp for slitless spectroscopy
  (source Par136 ID14 in G141 grism data). Blue symbols and labels
  correspond to primary target source; green color represent
  contaminants, secondary sources. {\it Top panel:}  \hh\ imaging of the
  sources. {\it Second panel:} original 2D stamp after aXe reduction. {\it Third panel:} 2D stamp after
removing the secondary sources with aXe. {\it Fourth panel:} 2D stamp after
removing the secondary sources with our models. {\it Fifth panel:} residual
after removing our 2D model of the primary source. At $\rm \sim 12 200\,\AA$ the residuals from an emission line are visible.}
   \label{fig:2Dspeclean}
\end{figure*}

One of aXe's features is the two-dimensional (2D) modeling of the
source's spatial profile for each galaxy. In what follows we refer to
the target galaxy as the primary, and to other galaxies which may
contaminate the spectrum as secondaries. In Figure~\ref{fig:2Dspeclean} we show a typical example. The second
panel of Figure~\ref{fig:2Dspeclean} shows the 2D dispersed stamp
extracted by aXe. The stamp is centered on the spectrum of the primary
(circle in the top panel), and the spectra of two secondary objects
are visible below it. Within aXe, source profiles for each
galaxy are fitted at each wavelength assuming a {\it single} Gaussian
profile. Although the single Gaussian is a sufficient approximation
for faint, poorly resolved sources, this assumption breaks down for
brighter resolved galaxies (such as those in the example shown in
Figure~\ref{fig:2Dspeclean}). Resolved, brighter sources typically
present extended wings, which are poorly modeled by a single Gaussian
profile.

After testing different functional forms to describe the extended
wings, we converged on describing each galaxy's profile
with a combination of two Gaussians sharing the same center, but with
different amplitudes and width. We fit a profile of the form:

\begin{equation} 
G_2(y)=A_1\cdot e^{\Big(\frac{y-Y_c}{S_1}\Big)^2}+A_2\cdot
e^{\Big(\frac{y-Y_c}{S_2}\Big)^2}; \label{eq:doubleG}
\end{equation} 

\noindent
where $A_1,$ and $A_2$ are the amplitudes of the two Gaussian
components, $S_1,$ and $S_2$ are the width, and $Y_c$ is the
coordinate of the center in the crossdispersion direction. This function provides the best compromise
between functional flexibility and number of parameters to adjust.  

As it is clear from Figure~\ref{fig:2Dspeclean}, the contamination
from secondary sources change with wavelength (e.g., due to relative
position of sources in the sky, spectral features like absorption
breaks and emission lines, variation of the grism transmission). For
this reason, we fit a combination of $G_2(y)$ functions (one for the
primary and one for ach secondary) to the spatial profile extracted
from the 2D dispersed stamps after a three pixel binning in the
wavelength direction.  Note we are only binning the data to perform
the fit. The spectral resolution for the rest of the analysis was not degraded

The fit is performed by minimizing the $\chi^2$ between the models and
the observed profile at each wavelength position. The first guess for
the parameters ($A_1,\,S_1,\,A_i,\,S_i$) are estimated from the direct
F110W images -for the G102 spectra- and F160W images -for the G141
spectra-. Contrary to their appearance in the dispersed images,
galaxies are typically well separated in the sky, allowing us to
obtain accurate relative positions and widths for all components
entering the fit.  

In Figure~\ref{fig:2Dspeclean} we show an example of this
procedure. The second, third and fourth pannels show an original (aXe
output) 2D dispersed stamp, a version clean from secondary
contaminants using the aXe model and a version cleaned using our
procedure, respectively. 

\begin{figure}[!t]
   \centering
   \includegraphics[width=8.5cm]{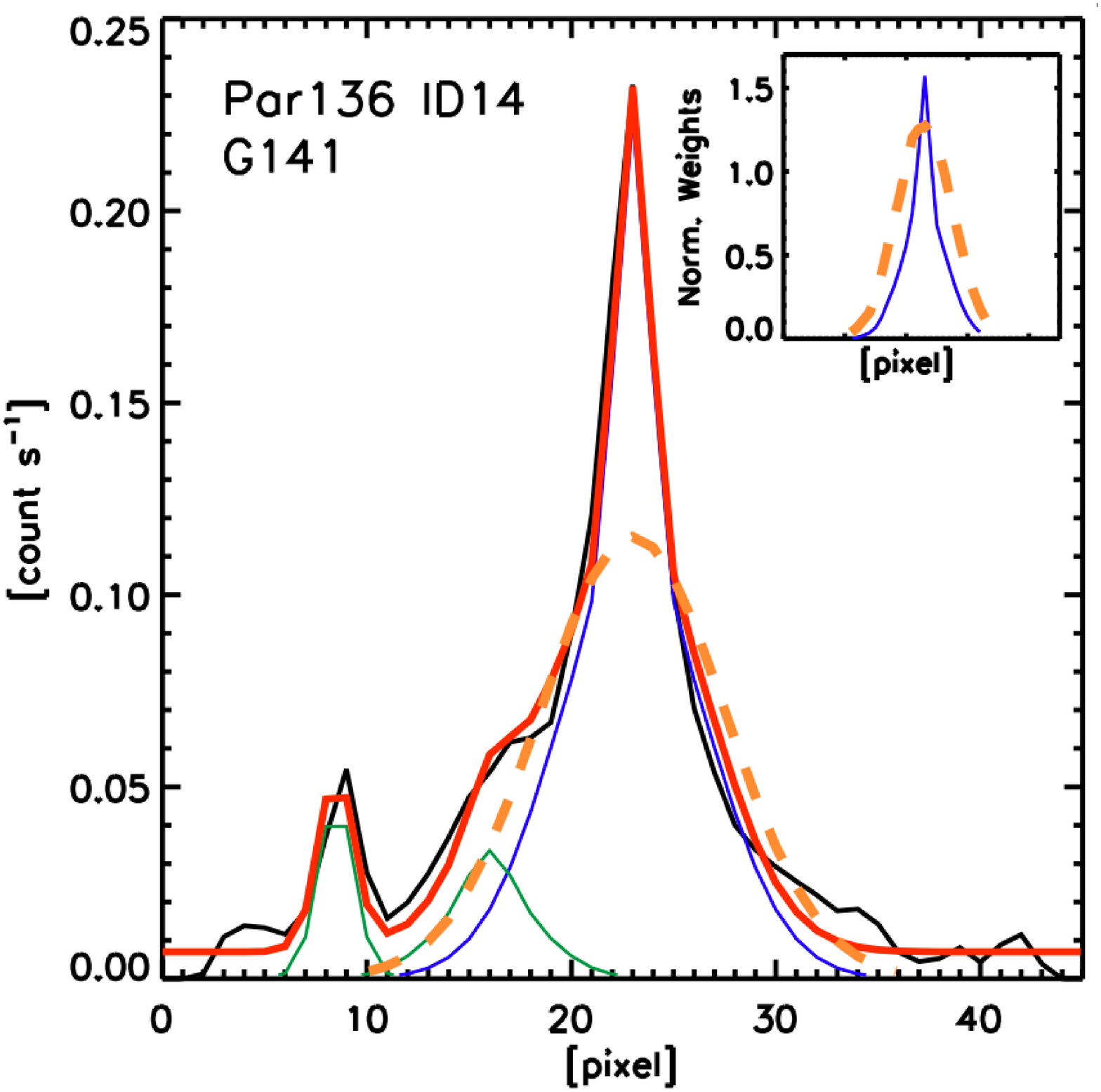}
  \includegraphics[width=8.5cm]{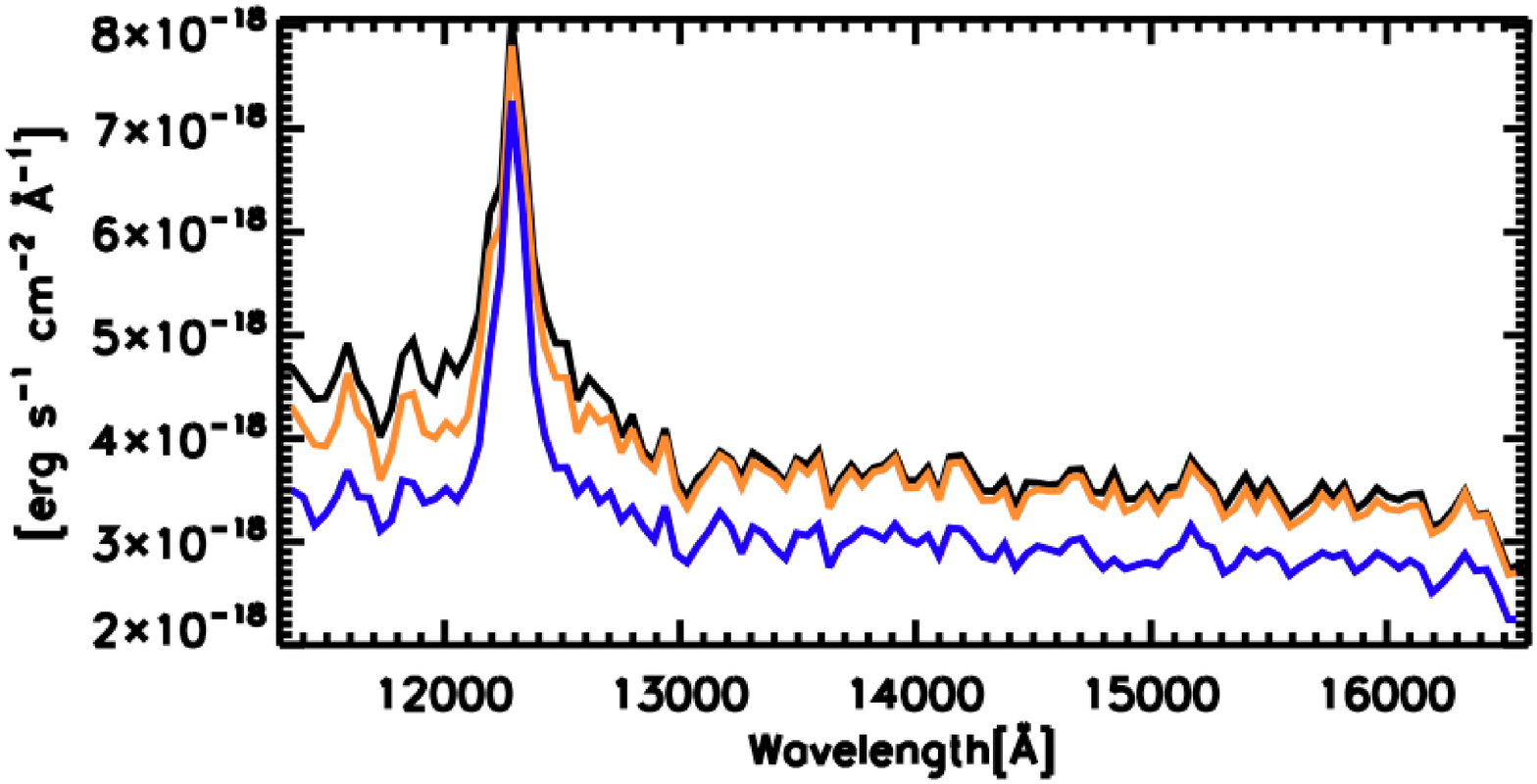}
  \caption{Example of mean 1D spatial profile and 1D spectral
    extraction (source Par69 ID12 in G102 grism data). {\it Left panel:} mean
  spatial profile of Par69 ID12 spectral stamp. Blue is the best model
to the primary source; green is the best model of the contaminant
secondary; black is the data and red, the best total model. The orange
dashed line is the mean primary profile from aXe. The inner panel in
the top right show aXe and our primary profiles with an arbitrary
normalization. This allows a fair comparison of the profiles as mean
weight masks. {\it Right panel:} Extracted 1D spectrum. In black, the data
is shown without cleaning. In orangeand blue we show aXe and our own extractions, respectively.}
   \label{fig:specleanprofile}
\end{figure}

In the top panel of Figure~\ref{fig:specleanprofile} we show the mean
spatial profile of Figure~\ref{fig:2Dspeclean} example target (black curve). We show
our best fit component to the primary (blue curve) and the secondary
(green curve) profiles, and also include the best fit aXe estimation
of the primary profile (dashed orange curve). In the inset
panel, we show aXe's and our profiles with a common (arbitrary)
normalization. Because in the optimal spectral extraction the profile
is used to weight each pixel in the final spectrum, the aXe profile
assigns relatively more weight to the lower $S/N$ wings with respect
to the peak flux. This effect, together with a poorer aXe removal of
contaminants, may translate into an overestimate of the total flux in
the 1D extracted spectrum, and in a lower $S/N$.

\begin{figure}[!h]
   \centering
   \includegraphics[width=8.5cm]{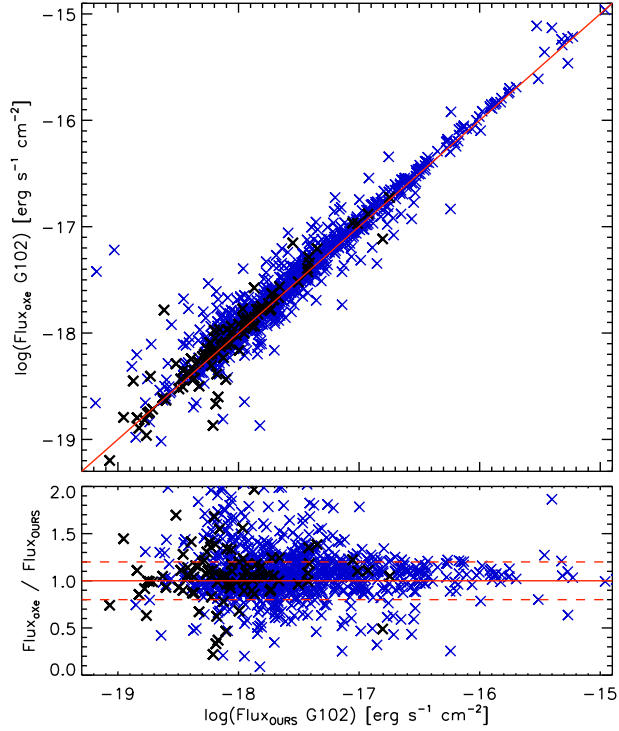}
\caption{In this comparison between aXe and our 1D spectral extractions we 
illustrate the differences between their integrated fluxes in G102 grism (8000-11250$\rm \AA$). 
aXe systematically overestimates the fluxes between 10-50\% from bright to faint galaxies. Blue symbols
  correspond to a \jj\ $\leq$23.0 selected sample from WISP
  fields. Black symbols correspond to the 41 galaxies used in this study
  (see Sec.\,\ref{results:sampleSPP}). Red continuous line is the 1:1 relation. Red dashed
  lines are the $\pm 20\%$ deviations from the 1:1 trend.}
  \label{fig:g102comp}
\end{figure}

In Figure~\ref{fig:g102comp} we compare the integrated flux in the
wavelength range 0.8--1.1\mic\ computed in aXe's and our extraction.
In the comparison, we consider only sources brighter than \jj$\le 23$.
Although there is a broad agreement between the 2 measurements, the
residuals (shown in the bottom panel) show that aXe's fluxes are
systematically overestimated from about 10\% for bright objects up to
$\approx 50$\% for some faint sources. The flux overestimate is due to
different effects for bright and faint sources. In fact, what
dominates in the bright sources is the poor fit of the single Gaussian
profile, while at the fainter level the dominant contribution is due
to the poor removal of nearby sources.

\section{Contamination from Stars}\label{app:rm_stars}
\begin{figure}[!h]
   \centering
  \includegraphics[width=8.5cm]{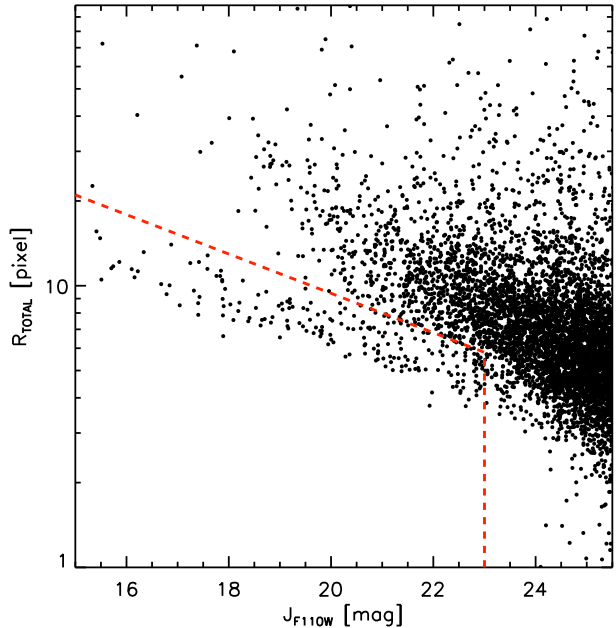}
  \caption{The size versus \jj\ diagram shows that contamination from stars (sources within the dashed red area) 
 can be removed as they define a distictive sequence with respect to galaxies. Black symbols are all \jj\ detections in
    photometry. The dashed red lines define a star-selection area for sources
    brighter than \jj$\leq 23$.}
  \label{fig:starrm}
\end{figure}

The contamination from foreground stars was removed using a
Magnitude-Size diagram. The $J_{F110W}$ measurements and total-light radius
estimations from Source Extractor software (\citet{bertin1996}) were used to separate galaxies from stars. In Figure\,\ref{fig:starrm}
we see how stars form a well defined trend. All these sources with
magnitudes $J_{F110W}\leq 23$ were flagged as stars
(similar magnitude limit than our final galaxy sample). Some residual contamination from stars
could be expected among the reminning sources (galaxies) and vice versa. We pay
special attention to this while selecting our final galaxy sample (see
Section\,\ref{sec:selectETG}).

A total of 234 stars were flagged
in this way. Another 71 stars (out of the demarcated area in
Figure\,\ref{fig:starrm}) were identified individualy while checking
individual galaxy spectra and photometry. Therefore, a total of 305
stars were flag and removed from our 27 fields.

\section{Effect of diluted emission lines in continuum flux}\label{app:emilines}
 
We have studied the effects of diluted emission lines in the continuum in order to establish if this has an important effect on the SED fitting and therefore in the retrieved SPPs. In all the SED fits presented in this paper, regions with detected emission lines have been masked to include only flux from the continuum. Potentially problematic cases, however, come from undetected emission lines (e.g., low S/N, large FWHM) that might artificially increase the continuum flux used in the SED fits. We explored these cases by modeling passive SEDs with emission lines. With these simulations we attempted to model the worst possible scenarios for the detection of emission lines. The models were made as follows:

\begin{itemize}

 \item We use the \citet{bc2003} model of a typical quenched galaxy ($\rm \tau=10\,Myr$, $\rm M_{\rm star} =10^{11} M_{\odot}$, SFR=0, solar metallicity, $\rm A_{V}=0$) and include emission lines represented as Gaussians. We model the lines [OIII]$\rm_{5007}$ and H$\alpha$ as those are the most common features detected in our data set. In a first step, a variety of EWs (between 35\,\AA\ -our detection limit- and 500\,\AA) and intrinsic FWHMs (between 40\,\AA\ -the pixel size on G141- and 300\,\AA) were explored for these lines. 

\item The spectra were redshifted to $z$=1.3 so the [OIII] and H$\alpha$ lines are both in G141, the poorer resolution grism. Then we degraded the spectra to G141 resolution. We considered the larger angular size of a galaxy in our sample along the dispersion direction of the slitless spectroscopy (525\,\AA\ in G141) to mimic the worst case resolution where emission lines can be easily missed.

\item In these simulations we use a noise range characteristic for our galaxy sample. The S/N per pixel typically ranges between 20 and 80 with a median of 40 (corresponding to a S/N per \AA\ of 0.8 in G141)

\item We produce a library of models with different combinations of S/N, line EW and FWHMs. Those cases where the emission lines are lost in the continuum were flagged. The task was performed by both, using the automatic line finder of Colbert et al.\ (2013, submitted), and by eye.

\item This resulted in a set of parameter ranges between which the emission lines are not detected (being diluted in the continuum): EW: 35-50\,\AA\ (above 50\,\AA\ the lines are always identified), FWHM: 50-270\,\AA\, S/N: 30-80 (below 30 the lines can not be distinguish from the noise).

\end{itemize}

\begin{figure}[!t]
   \centering
   \includegraphics[width=10cm]{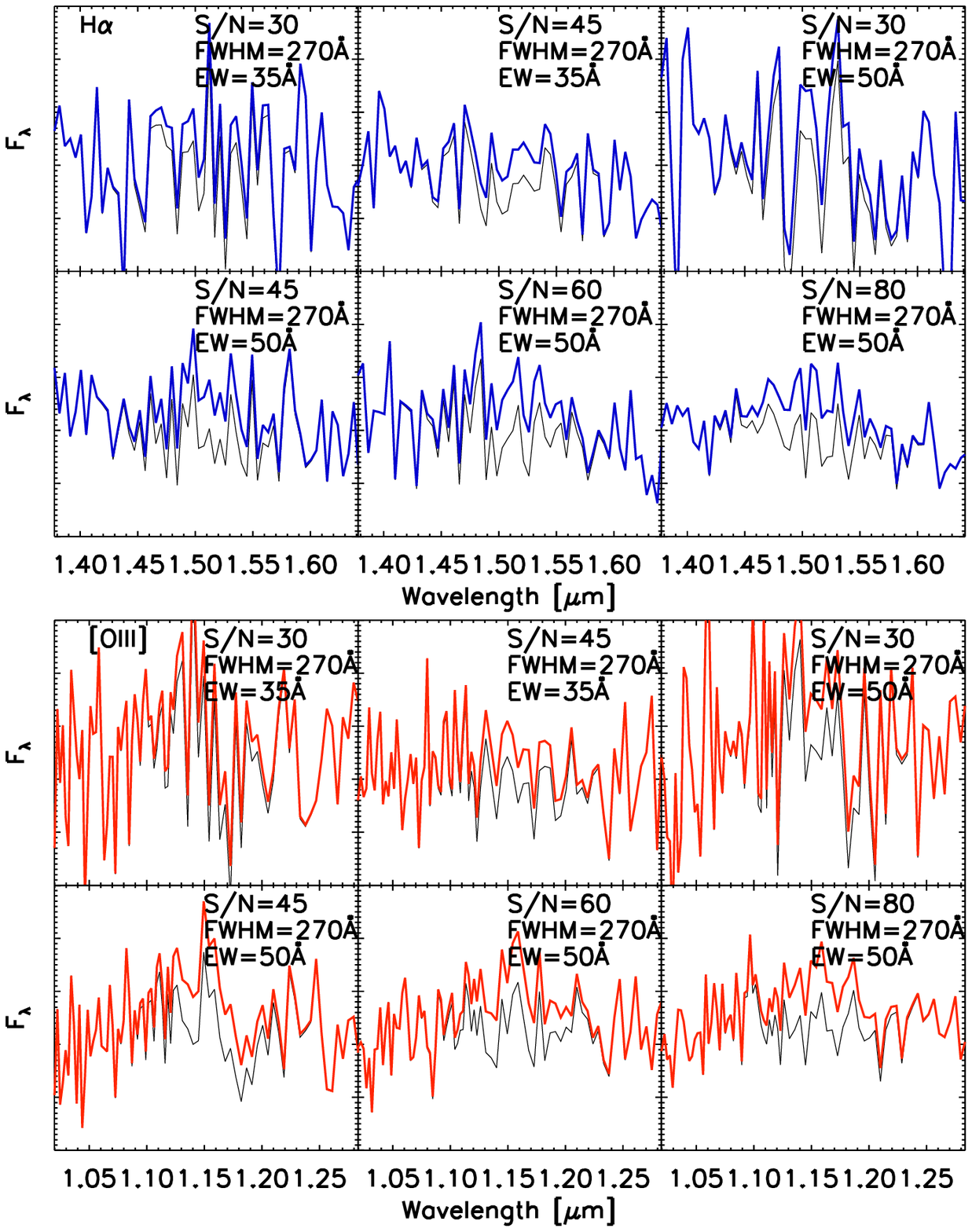}
   \caption{Model examples of combinations of S/N, EW and FWHM that cover the parameter space where emission lines are not detected. Top panels show H$\alpha$. Bottom panels, [OIII]. In black, continuum spectra; in colors, continuum+emission line.}
   \label{fig:emiexample}
\end{figure}

In Figure\,\ref{fig:emiexample}, we show examples of 6 combinations of S/N, EW and FWHM that cover the parameter space where the lines are not detected (6 for H$\alpha$ and 6 for [OIII]). In black, the pure continuum spectra while in color, spectra+emission line. 

\begin{figure}[!t]
   \centering
   \includegraphics[width=8.5cm]{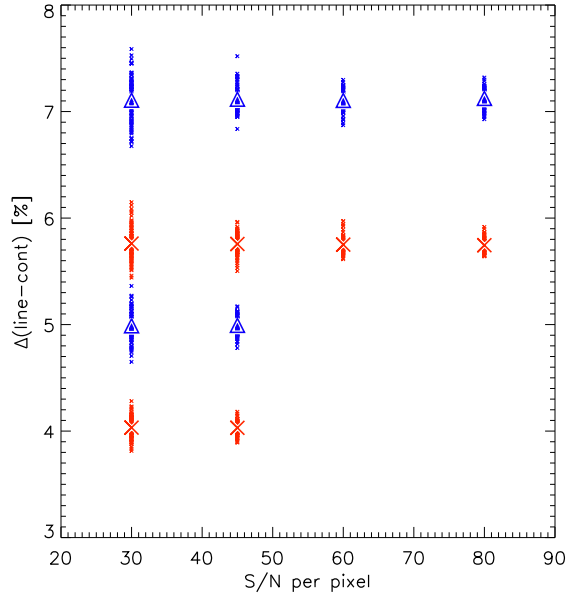}
   \caption{The percentage difference between fluxes in the model continuum and continuum + emission spectra are $< 7\%$ for all S/N (H$\alpha$ in blue, [OIII] in red). Small 'x' symbols are 100 MC simulations for each case while large symbols are the mean values.}
   \label{fig:emiflxpercent}
\end{figure}

For the 6 combinations of parameters we run 100 MC simulations with different random noise. Then we calculate the mean flux within 3-$\sigma$ of the emission line in both, the pure continuum and continuum+emission spectra. Finally, we estimate the percentage increase in flux in the emission-line cases with respect their pure continuum spectra. In Figure\,\ref{fig:emiflxpercent} we show those percentages versus S/N (H$\alpha$ in blue, [OIII] in red). Large symbols are the means while the small '$\times$' symbols are the 100 MC for each case.

As we see the differences in flux are $\lesssim 7\%$, which is below the $>15\%$ flux uncertainty in the grism spectroscopy. Therefore, in those cases were emission lines are diluted in the continuum, the flux contribution from the line is lower than our 1-$\sigma$ errors in the grism data. This implies that as far as we mask the detected emission lines for our SED fits, there is no need to make specific modeling of emission lines for our SED fitting process and simulations.

\section{Systematic uncertainties in redshift and stellar population parameters}\label{app:systerr}

In this appendix we present a study on the systematic effects of SED fitting in our data.  Our aim is to constrain for which specific data sets (e.g., with UVIS detections or upper limits, with or without IRAC) and SPP-ranges our predictions are reliable, also providing systematic errors for the different SPPs. Our study is based on simulating our data set at different redshifts with stellar population synthesis models to later recover the (known) redshifts and SPPs of these models using our customize {\tt IDL} $\chi^2$-minimization code (Section\,\ref{sec:redshift} and \ref{sec:sedfit}).

We simulate our photometric and spectroscopic data by using \citet{bc2003} models. The complete single-burst model library considered to retrieve SPPs of our galaxy sample was used here (see Section\,\ref{sec:sedfit} for the different SPP-ranges considered). At this point we want to mention that assuming the SFHs of our sample galaxies have a single-burst component is a simplification. Real galaxies may (and probably do) have more complex SFHs, like those that include multiple star-bursts during their lives. A proper exploration of these scenarios for the estimation of systematic errors, however, would substantially increase the number of models to be tested. Among others, multiple combinations of different numbers of star-burst, occurring at different times, with different intensities and $e$-folding time-scales would be needed. In literature, such number of simulations are not usually (if ever being) attempted for the kind of analysis presented in this appendix, mainly because the immense amount of computational time is prohibitive. Therefore, we decided to limit our simulations to the original single-burst model library. As a consequence we remind the reader our systematic error estimates should be consider as lower-limits.

The model SEDs were re-scaled to agree with typical data broad-band fluxes and corresponding stellar masses. We multiplied all the SEDs for a given factor such that a typical galaxy in our sample (exponentially declining SFH with $\rm \tau=10\,Myr$, 2\,Gyr old, no star-formation and no extinction) would have a stellar mass of $\rm 10^{11}M_{\odot}$. Variations in the stellar mass between 10.6 and 11.8\,dex (the range of our data) do not produce any significant change in the results of our simulations. Therefore most of the results presented here correspond to an input model mass of $\rm \approx 10^{11}M_{\odot}$ (a mild spread in input mass, typically of $\rm \pm 1\times 10^{10}M_{\odot}$, occurs given the different combinations of model SPPs).

 Redshifts between 0.6 and 2.4 in steps of 0.2 were tested and the model SEDs were redshifted accordingly. Using a wide redshift range for the models allowed us to explore the effects of redshift misidentification in the error budget and derived quantities. For $z<0.6$ none of the models pass our \jj$-$\hh$\geq 0.6$ galaxy selection criterion. At $z > 2.4$ the contribution to the mass function integral is very small \citep{marchesini2009}. 

Then we degraded the models to G102 and G141 spectral resolutions to obtain our simulated grism spectroscopy. Random noise was also added to these models in order to match our data. A signal-to-noise per pixel $\sim 40$ was adequate to represent most of our galaxy spectra. Then the model spectra were binned in wavelength to reproduce the slitless-resolution dependency on the angular size of the source in the dispersion direction. For the angular size we use the median luminosity-profile-FWHM of our sample galaxies in \jj-band, corresponding to $\rm 5\,pixels$ in the WFC3 detector ($125$ and $\rm 237$\AA\ in G102 and G141, respectively). 

In addition to simulating the grism spectra, we use the model SEDs to simulate broad-band photometry (F475X, F600LP, F110W and F160W filters). For the two WFC3/UVIS bands we considered a typical data $1\,\sigma$ uncertainty of 0.15 mag. Their fluxes were randomized using a Gaussian distribution with this width. Then we compared the model UVIS magnitudes with typical $3\,\sigma$ upper limits from our galaxy master sample (25.4 and 24.9 mag in F475X and F600LP, respectively). If a model UVIS magnitude was brighter than the limit, it was considered as a detection with its corresponding uncertainty. If equal or below the limit, the UVIS magnitude was assigned the $3\,\sigma$ upper limit for the band and considered as such for the $\chi^2$-minimization process. We measured and randomized the IRAC $3.6\,\mu$m photometry in an analogous way with a characteristic error of 0.3 mag.

Once we had our SED library redshifted and simulating real data conditions, all the models passed through the same selection criteria in color (\jj$-$\hh $\geq 0.6$) and magnitude (\hh $\le 23.0$) than our galaxies. Then we used our customized {\tt IDL} code to recover redshifts and SPPs of those models that passed the selection. In what follows, the original redshifts and SPPs from which the model SEDs were made are referred as ``input parameters'' (or sub-index ``IN'', e.g., $z_{\rm IN}$, Age$_{\rm IN}$) while the redshifts and SPPs retrieved by our {\tt IDL} code are referred as ``recovered parameters'' (or sub-index ``OUT'', e.g., $z_{\rm OUT}$, Age$_{\rm OUT}$).

\begin{figure}[!h]
   \centering
  \includegraphics[width=8.5cm]{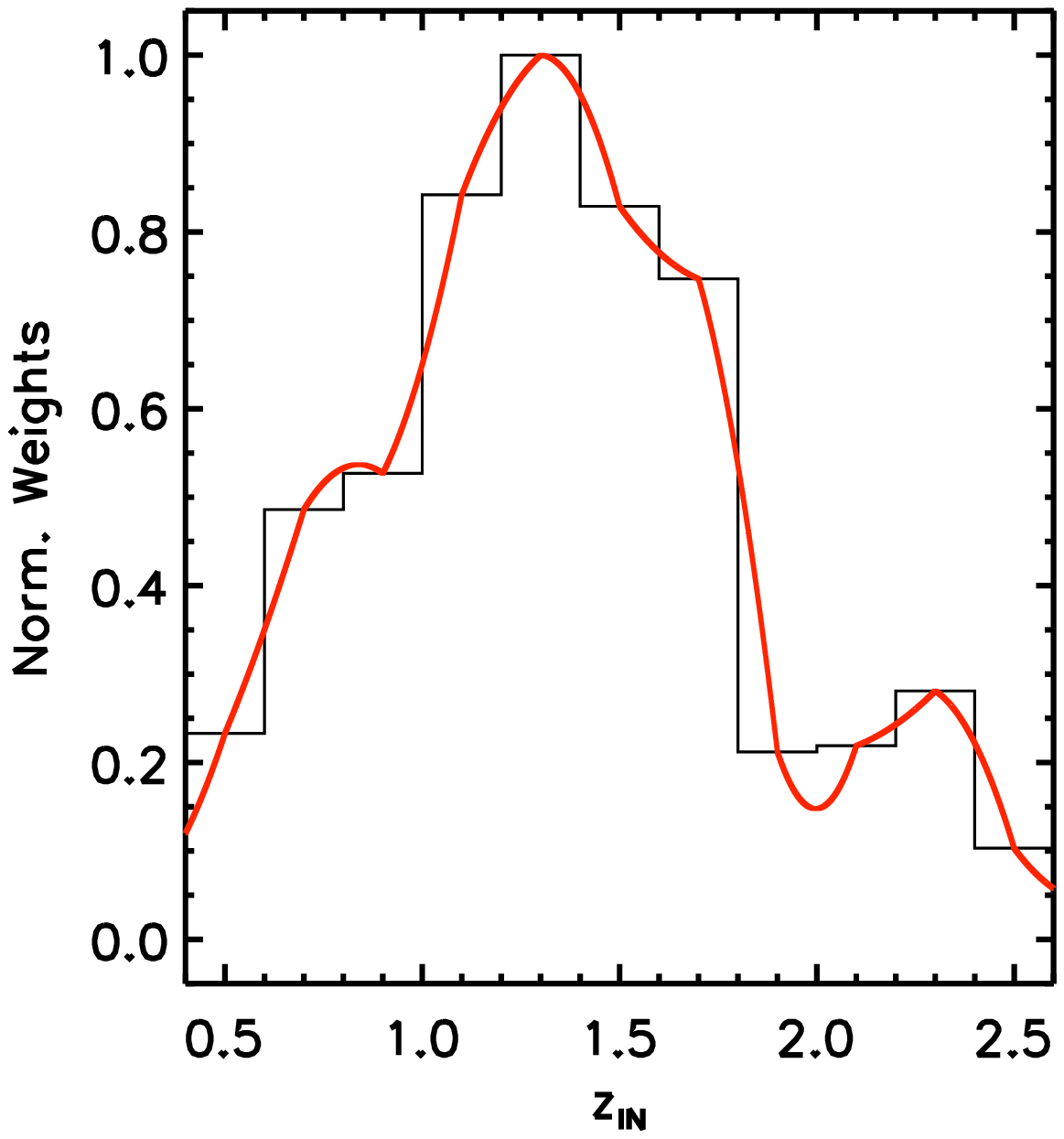}
   \includegraphics[width=8.5cm]{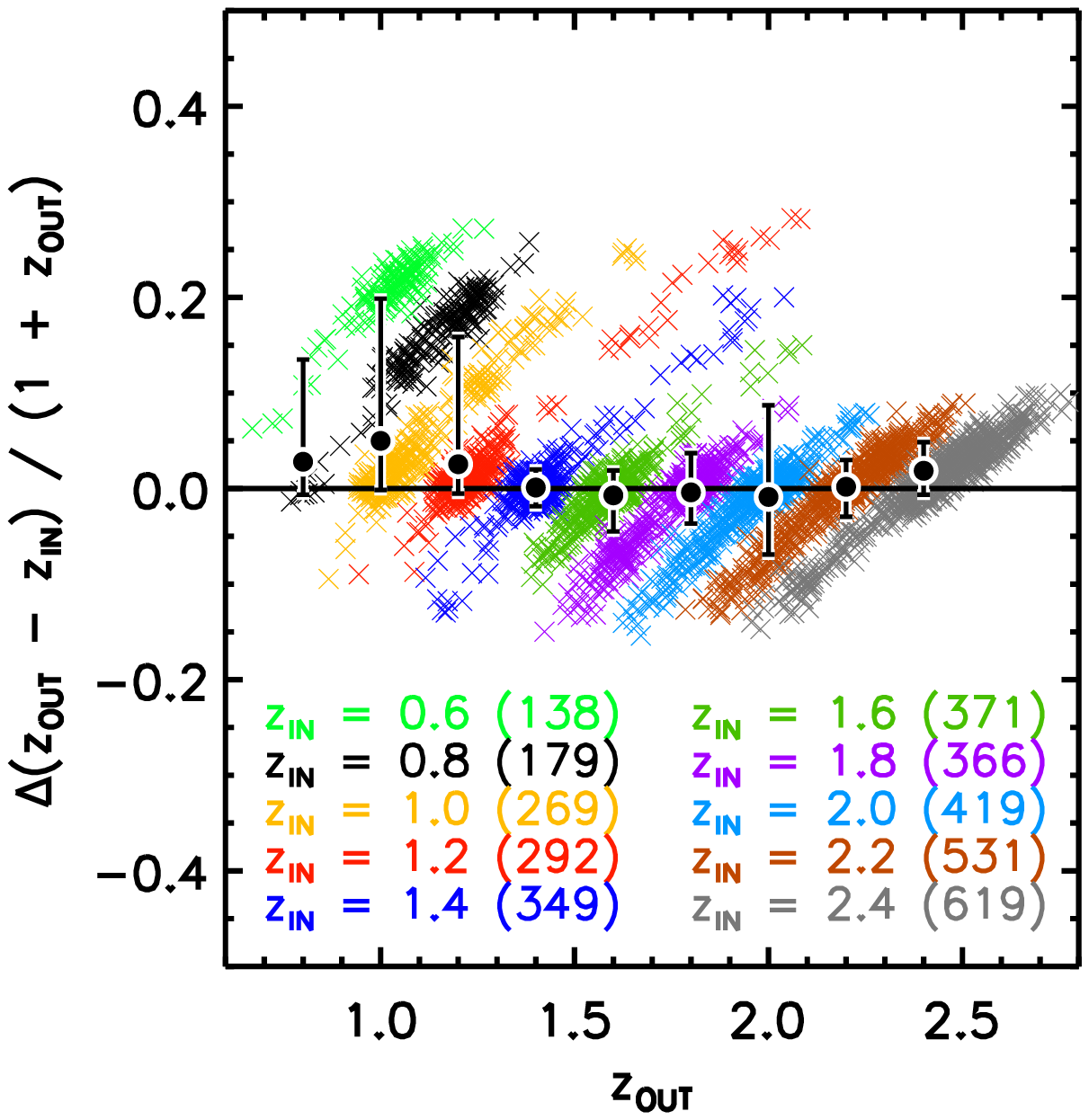}
  \caption{{\it Left:} Peak-normalized histogram of weights for retrieved redshifts and SPPs as function of $z_{\rm IN}$ (black). In red, the smoother function used for our calculations. The weights are from the redshift distribution of CANDELS \citep{candels} after selecting galaxies with the same color-magnitude criteria used in our data. {\it Left:} Systematic error in $z_{\rm OUT}$ versus $z_{\rm OUT}$. Weighted mean systematic offsets for nine bins (0.1 wide in $z_{\rm OUT}$) are shown as black dots. Error bars represent the 68\% percentiles. In the bottom we show the color code for $z_{\rm IN}$ and in parentesis the number of models that pass our color-magnitude selection criteria. The positions of individual data points have been slightly randomized for presentation purposes.}
   \label{fig:z_out}
\end{figure}

In order to make a reliable representation of our data set we need more than retrieving the galaxy SPPs and modeling the specific features of the photometric and spectroscopic data. When observations cover a given redshift range, different selection effects are present like the different cosmic volumes and regions of the luminosity (mass) function sampled at different redshifts. We took into account these effects in our systematic error calculations by assign a 'relative weight' to each simulated SED as function of redshift ($z_{\rm IN}$). The selection effect was introduced as a normalized multiplicative factor for a retrieved redshift or SPP (e.g., $z_{\rm OUT}$, Age$_{\rm OUT}$). The weights come from the redshift distribution of galaxies from CANDELS \citep{candels} after selecting sources with the same color-magnitude criteria used in our data. A normalized histogram of the redshift distribution  is shown in the left panel of Figure\,\ref{fig:z_out}, together with a smoothed version used in our calculations.

In the right panel of Figure\,\ref{fig:z_out} we show the systematic errors in $z_{\rm OUT}$ versus $z_{\rm OUT}$ with median systematic offsets for nine redshift bins (black dots) and their 68\% percentiles. In this plot we clearly see that for $z_{\rm IN} \leq 1.0$ the redshifts are completely misrecovered having in most cases values $z_{\rm OUT} \geq 1.0$. The fact we were not sampling the 4000\AA\ break with the spectroscopy at low-$z$ has a relevant systematic effect for $z_{\rm OUT} \leq 1.2$. As a result of this redshift-recovering tests we decided to remove from our galaxy sample all sources with recovered spectrophotometric redshifts $<1.0$ as they are not reliable. For redshifts $\geq 1.0$ the systematics are consistent with zero, within the uncertainties. We present those uncertainties in Table~\ref{tab:systerr}. 

As retrieving spectrophotometric redshifts is the first step in characterizing our galaxy sample, any uncertainty in the systematic error of this parameter will have repercussions in the recovered SPPs. For this reason the systematics for most of the SPPs shown in Table~\ref{tab:systerr} are also presented at different redshift ranges.

\begin{figure}[!t]
   \centering
  \includegraphics[width=18cm]{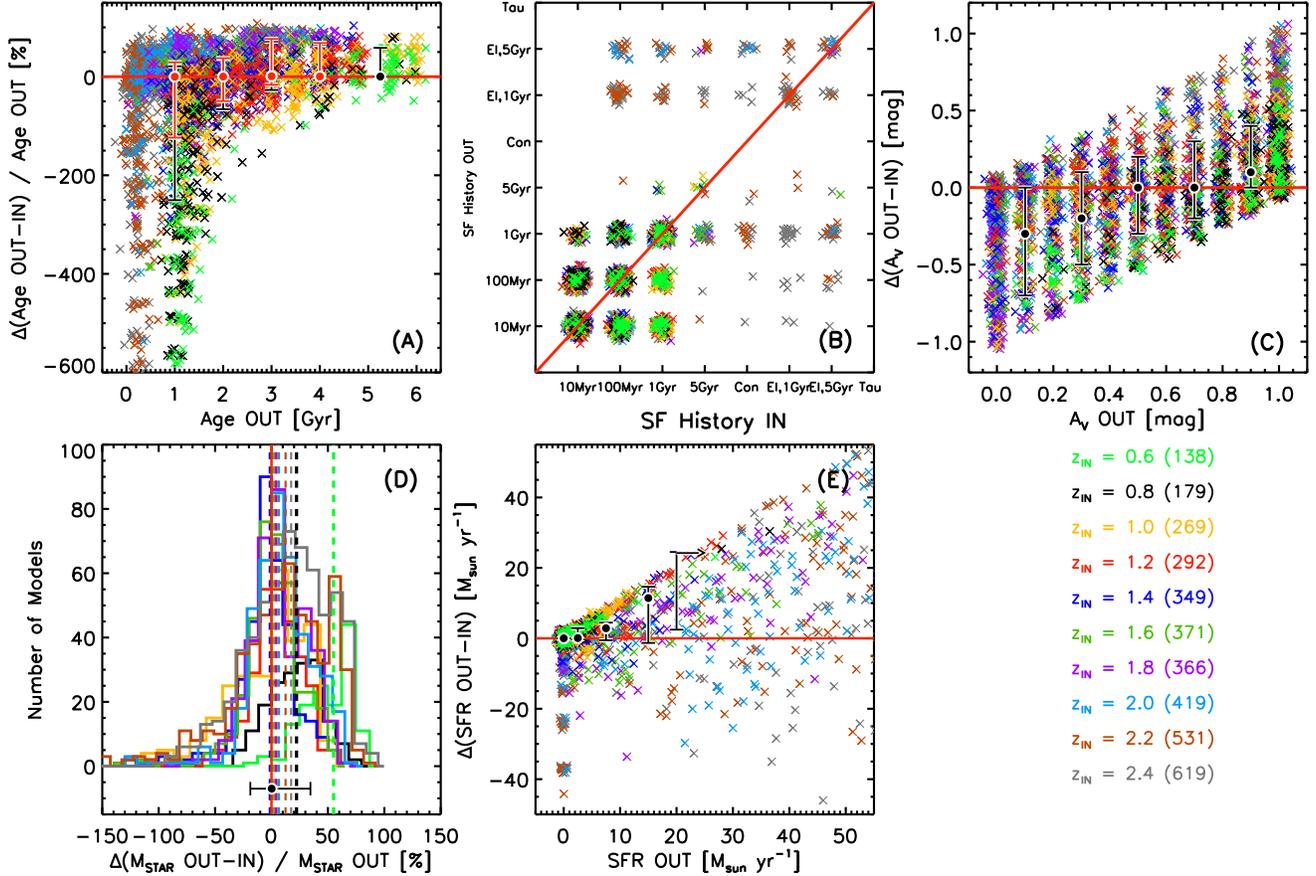}
 \caption{ Comparison between input and output SPPs for $0.6 \leq  z_{\rm IN} \leq 2.4$ models and stellar mass $\rm \sim 10^{11}M_{\odot}$. In panel (A) we show the percentage difference between input and output ages versus $\rm Age_{OUT}$. Black dots with error bars represent weighted mean (\%) systematic offsets and 68\%-percentiles in bins of 1\,Gyr in $\rm Age_{OUT}$. The red dots and error bars only consider models with $z_{\rm OUT} \geq 1.3$. In panel (B) we show $\rm SFH_{OUT}$ vs $\rm SFH_{IN}$. In (C) we present  the difference between input and output extinctions versus $\rm Av_{OUT}$. Dots with error bars represent weighted mean systematic offsets and 68\%-percentiles in bins of 0.2\,mag. In panel (D) we show histograms for recovered stellar mass ($\rm \sim 10^{11}\,M_{\odot}$) at different redshifts. Vertical dashed lines represent the median of each distribution. The black dot with error bars represents the weighted mean (\%) systematic offset in stellar mass. In (E) we present the difference between input and output SFRs versus $\rm SFR_{OUT}$. Dots with error bars represent weighted mean systematic offsets and 68\%-percentiles in different bins. The last dot with an arrow represent the systematic offset for $\rm SFR_{OUT}>20\,M_{\odot} yr^{-1}$. The different colors in 'X' symbols and histograms represent different $z_{\rm IN}$ as described in the leyend (number of models per $z_{\rm IN}$ in parenthesis).}
   \label{fig:inout_all}
\end{figure}

In Figure~\ref{fig:inout_all} we show comparisons between input and output Age, SFH, extinction (A$_V$), stellar mass and SFR for $0.6 \leq  z_{\rm IN} \leq 2.4$ models. The different colors in symbols and histograms represent different $z_{\rm IN}$ as described in the legend. The continuous red lines indicate no difference between input and output values.

In panel (A) of Figure~\ref{fig:inout_all} we show the percentage difference between input and output ages ($\rm \Delta age$) versus $\rm Age_{OUT}$. The median systematic offsets (black dots) clearly show large uncertainties for ages $\rm \sim 1\,Gyr$. We notice, however, that the larger contribution to these offsets comes from low-$z_{\rm IN}$ were the 4000\AA\ break is not sample with the spectroscopy. On the other hand, we found an important degeneracy between age and $\rm A_{V}$. This degeneracy also contributes to the systematics in $\rm Age_{OUT}$-$\rm \Delta age$ described before, and its strength strongly depends on how well can we constraint these SPPs at different redshifts. The $\rm \Delta age$ versus $\rm Age_{OUT}$ systematics become more evident at lower redshifts where we lack the spectroscopic constraint of the 4000\AA\ break. But even at larger redshifts we also observe some degree of degeneracy between young-underestimated ages and high-overestimated extinctions. Taking all this in consideration we decided to proceed as follows: we only considered ages at $z_{\rm OUT} \geq 1.3$ as constrained enough in order to be discussed in this paper. The extinctions and lower-redshift age estimations are too degenerate to make meaningful statements based on them. 
As for the redshifts, the systematics in $\rm Age_{OUT}$ are consistent with zero within the uncertainties. We show these  results in Table~\ref{tab:systerr}.

In panel (D) of Figure\,\ref{fig:inout_all} we show histograms for the difference (in \%) between recovered and input stellar masses. In this particular case we show models with $\rm M_{star} \approx 10^{11}\,M_{\odot}$ as this is a representative mass for our galaxy sample and there are no strong variantions in the results within our galaxy mass range. The stellar mass is the most robust SPP retrieved with our data. Between $1.2 \leq z_{\rm OUT} \leq 1.8$ there are virtualy no systematic offsets within $\rm log \,(M_{STAR,OUT}/M_{\odot}) = 10.6$-$11.8$ ($\leq 10\,\%$). At  $z_{\rm OUT} < 1.2$ systematic offsets become more relevant, varying with stellar mass from 30, 20 and 10\% at $\rm log\, (M_{star}/M_{\odot}) = 11.8$, 11.0 and 10.6, respectively (on average). These systematics are shown in Table~\ref{tab:systerr}.

For the remaining two SPPs (SFR and SFH) different combinations of data sets could determined if the recovered values are (or not) reliable. In panel (B) of Figure\,\ref{fig:inout_all} we show output versus input SFHs. From this figure we highlight two points. First, for our entire redshift range we can distinguish between short ($\tau \leq 100$\,Myr) and long ($\tau \geq 1$\,Gyr) SFHs, although we cannot separate SFHs with $\tau = 100$\,Myr and $\tau = 10$\,Myr. Second, short SFHs ($\tau \leq 100$\,Myr) are recovered independently of the data set used to derive redshifts and SPPs. For more extended SFHs ($\rm \tau \geq 1\,Gyr$), however, we found that we are able to recover similarly extended SFHs only for data sets with 2 UVIS detections independently of the redshift. Overall, these simulations imply we were able to broadly discriminate between $\tau$ above and below $\rm 1\,Gyr$.

In panel (E) of Figure\,\ref{fig:inout_all} we present the difference between output and input SFRs versus $\rm SFR_{OUT}$. Over the entire redshift range we can successfully recover $\rm SFR_{OUT} \leq 10\,M_{\odot} yr^{-1}$ with uncertainties $\rm \leq 3\,M_{\odot} yr^{-1}$. For $\rm SFR_{OUT} > 10\,M_{\odot} yr^{-1}$, however, the uncertainties in the systematics increase steadily reaching $\rm 20\,M_{\odot} yr^{-1}$ at $\rm SFR_{OUT} > 20\,M_{\odot} yr^{-1}$. The different retrieved redshifts make no much of a difference on this trend, but only increase the scatter on the median systematic errors at high $z_{\rm OUT}$. For this reason, in Table~\ref{tab:systerr} we present the systematics in SFR independently of redshift.

The strong systematic over-prediction of the SFR for $\rm SFR_{OUT} > 20\,M_{\odot} yr^{-1}$, however, does not have a significant impact in our paper. First, we base the science on a quenched galaxy sample which always has the low-SFRs we can successfully retrieve. Second, we do not make claims based on the precise values of the SFRs. We only use it through the {\it specific} SFR to identify quenched galaxies ($\rm SSFR < 10^{-2}\,Gyr^{-1}$). As no further claims are made based on SSFRs, the relevant test is to determine if an intrinsically quenched/SF galaxy can be recovered as such with our procedure, independently of the exact value of their SSFR.

Stellar population properties like SSFR and SFH are closely linked. For example, at the time of observation a galaxy with an extended SFH is more likely to show a higher SSFR than a short-burst galaxy. Also, our ability to constraint these properties is tightly related to the constraints we can provide with the data. Of particular importance is the number of UVIS detections available to constraint these SPPs. Detections in the rest-frame UV are more likely to occur in SF galaxies. For these reasons we decided to explore the success in recovering SSFRs and SFHs for different UVIS data sets in our entire redshift range. In this exercise we did not explore individual values of SSFR and SFH but {\it ranges} within which these parameters are reliable. For the SSFR two regimes were defined as above/below $\rm 10^{-2}\,Gyr^{-1}$ (our quenched/SF selection criterion). As already mentioned in previous paragraphs, SFHs are defined as short-SFHs if $\rm \tau < 1\,Gyr$, and extended if $\rm \tau \geq 1\,Gyr$.

Our results show that the reliability of the recovered SFHs and SSFRs depends on the constraints from the UVIS data. At all redshifts we reliably recovered the quenched and short-SFH models (the relevant sources for this paper). This was independent on the level of constraint from UVIS data (two upper limits, one and two detections).  On the contrary, and just to give the complete picture to the reader, SF and extended-SFH models were harder to constraint. Only those models with 2 UVIS detections provided reliably recovered SFHs and SSFRs. \\

\begin{figure}[!t]
   \centering
  \includegraphics[width=18cm]{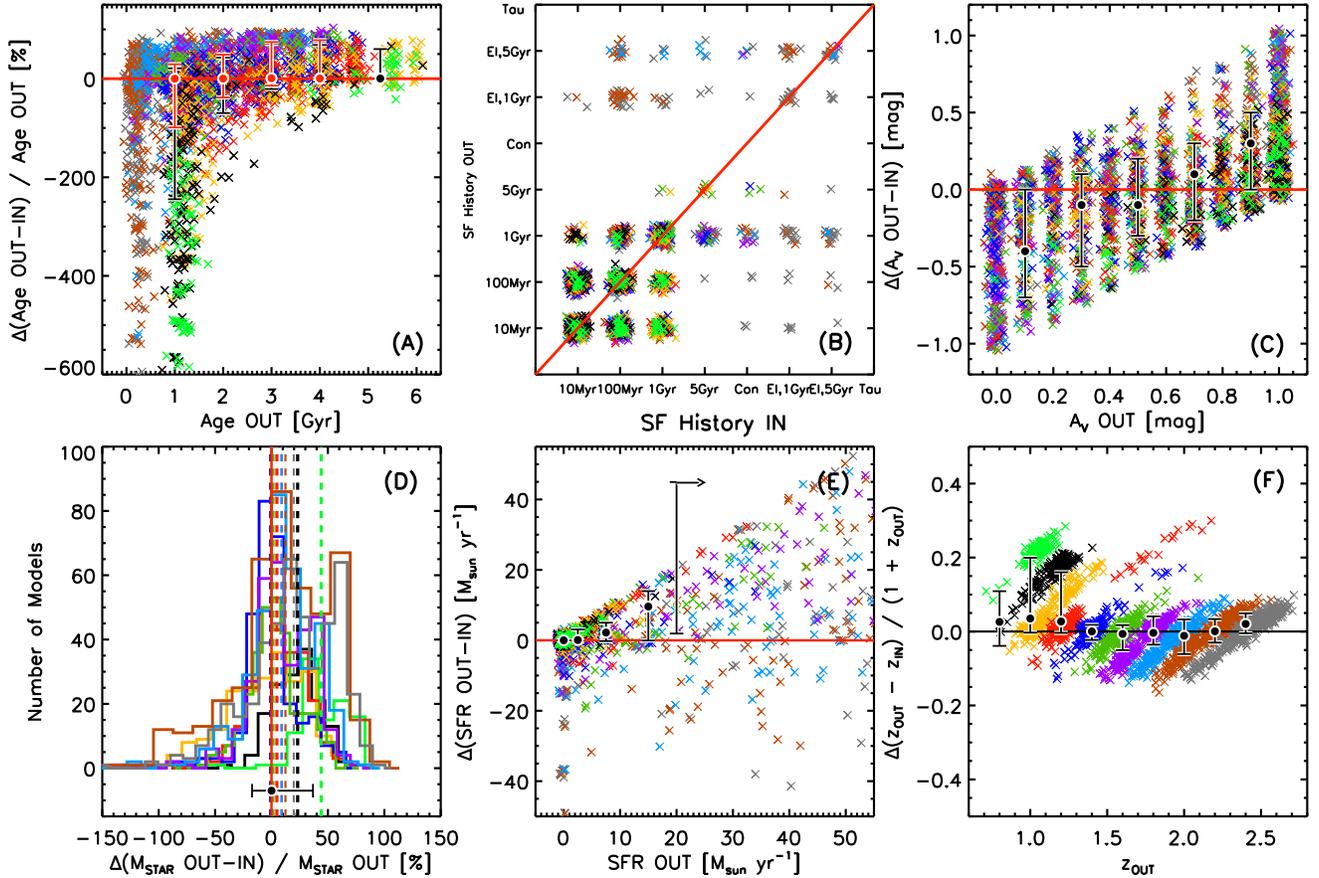}
 \caption{Same as Figure~\ref{fig:inout_all} but for SED fits without IRAC~$3.6\,\mu m$ data.  Analogously to Figure~\ref{fig:z_out} (right panel), in panel F we show the weighted mean systematic uncertainties in $z_{\rm OUT}$ for SED fits without IRAC data. The $z_{\rm IN}$-color code is the same as in Figures~\ref{fig:z_out} and \ref{fig:inout_all}.}
   \label{fig:inout_all_NOIRAC}
\end{figure}

In Figure~\ref{fig:inout_all_NOIRAC} we present the systematic uncertainties of all our SPPs and redshifts for SED fits on which no IRAC~$\rm 3.6\,\mu m$ data was used (referred as 'no-IRAC' from now on). A summary of the systematic errors for no-IRAC models is shown in the right column of Table\,\ref{tab:systerr}. We observe that $\sim 15\,\%$ of the model SEDs experience convergence issues during the $\chi^2$-minimization. This percentage is fairly constant for our entire range of $z_{\rm IN}$ models. In a closer analysis of these cases we found that models with $\rm A_{V, IN}>0.8\,mag$ are frequent among the nonconverging cases, particularly at $z_{\rm IN}<1.0$. This does not imply, however, that a majority of high-$\rm A_{V, IN}$ models do not converge. During the selection of our galaxy sample, the cases of nonconvergence for no-IRAC galaxies were very few. In any case, given the number of no-IRAC galaxies in our final sample, a $15\%$ of failed convergence would imply that at the most we are loosing 1 galaxy for this reason. Overall, we conclude from this analysis that the  IRAC~$\rm 3.6\,\mu m$ data makes it eaier to constraint sources with high extinction.

As a test, to avoid misinterpreting the slightly smaller number statistics of no-IRAC simulations with respect the IRAC counterpart, we remove from the later the non-convergent models of the former. We found that the systematic errors and dispersions of the simulations with IRAC barely change after removing those models. Therefore we can confidently interpret differences among IRAC / no-IRAC models as the consequence of lacking the $\rm 3.6\,\mu m$ data.

For all the SPPs no major changes were observed in the median systematic errors by using or not IRAC data. For $z_{\rm OUT}$ no changes in the systematics was observed with respect the simulations with IRAC. The systematics in stellar ages remain fairly stable, with a 10\% increment in the systematic uncertainties for $\rm z_{\rm OUT} > 1.8$. Only a reduction from 75 to 65\% appears for the younger ages at $\rm 1.3 \leq {\it z_{\rm OUT}} \leq 1.8$. However, as the uncertainties for that age regime are very large, we decided to keep 75\% of error in the systematics for the no-IRAC models for those age and redshift ranges. 

The SFHs are still well recovered within the broad categories defined above (exponentially declined with $\rm \tau \leq 100\,Myr$ and $\rm \geq 1\,Gyr$). The no-IRAC model runs only suggest a larger fraction of misidentified $\rm \tau_{IN}=5\,Gyr $ as $\rm \tau_{OUT}=1\,Gyr $. Also, without IRAC, less models are misclassified as exponentially-increasing SFHs.

The lack of IRAC data has limited effect on the median systematics of stellar mass. It mostly widens the distributions shown on panel D of  Figure~\ref{fig:inout_all_NOIRAC} were the stronger effects are observed at high-$z_{\rm IN}$. This implies that the stellar mass of a galaxy is probably better constraint by using IRAC, particularly at high retrieved redshift. However, this does not translate into a larger {\it average } offset in the systematics on no-IRAC models: only at $z_{\rm OUT}\geq 1.8$ a mild increment of 5\% in the systematic error is measured.  

The $\rm SFR_{\rm OUT}$ suffer from minor variations in the uncertainty on the systematics for no-IRAC models. 
As it might be expected at this point, the lack of IRAC data does not prevent us to constraint the SSFR as we did for the models with IRAC photometry.
\\

In summary, our study of the systematic uncertainties in the SED fitting procedure allowed us to identify those specific data sets (combinations of photometry and spectroscopy) which did not provide enough constraint to reliably recover certain SPPs. Apart of A$_V$ and ages from sources at $z$<1.3, quenched-galaxy SPPs are fairly well constrained independently of the data set used. Concerning the actual systematic uncertainties we retrieved, there is no significant bias in the derived stellar masses, (S)SFR, SFH and ages (the later at $z$>1.3) for galaxies with and without IRAC photometry. Our ability to constraint SPPs with our current data set is in part constrained by our uncertainty in redshift determination, for which the 4000\AA-break is the key ingredient. A summary of the systematic errors for each discussed SPP is presented in Table\,\ref{tab:systerr}.\\\\

Some final remarks: We should mention that systematic uncertainties would increase/decrease in case of an IMF variation within our galaxy sample. The variation in the IMF is a very debated topic where most of the recent developments are restricted to local galaxy data. A recent paper of \citet{cappellari2012} on local early-type galaxies suggests the existence of a range of IMFs from 'heavier-Salpeter' to Chabrier cases. In this work, however, and for the specific mass-range of our galaxy sample (assuming $\rm log(M_{star}/M_{\odot}) \geq 10.65$ galaxies have velocity dispersions $\rm \geq 200\,km\,s^{-1}$), there is no correlation between IMF and stellar mass. Galaxies in this mass-range spread from heavier-Salpeter to Chabrier like IMFs \citep[see also][]{spiniello2013}. \citet{mitchell2013} present a theoretical study on the systematic effects of SED fitting in stellar mass estimations. For different stellar-population-synthesis models (a Bruzual \& Charlot version and Maraston et al.\,2005) and IMFs \citep[][and a non-Universal IMF]{kennicutt1983} these authors find no IMF-dependency on mass estimations in our mass-range. For high masses the scatter in the systematics is large independently of model/IMF used with dispersions ranging from 20-300\% the value of the original input mass. All these results may imply that IMF variations are a galaxy-to-galaxy phenomenon, information we certainly lack. Therefore we chose to use a unique Salpeter IMF for our calculations as it seems to be 'representative' of the galaxy population we are studying. It also allowed us to directly compare our results with similar stellar population studies in literature.

\end{document}